\documentclass[a4paper,11pt]{article}
\usepackage{jheppub}
\pdfoutput=1
\usepackage{jheppub}
\usepackage[T1]{fontenc}
\usepackage{caption}
\usepackage{graphicx}
\usepackage{amsmath}
\usepackage{amsmath}
\usepackage{xspace}
\usepackage{subfig}
\usepackage{color}
\usepackage[utf8]{inputenc}
\usepackage{slashed}
\usepackage{fancyvrb}
\usepackage[utf8]{inputenc}
\usepackage[english]{babel}
\usepackage{amsfonts}
\usepackage{amssymb}
\usepackage{color}
\usepackage{cancel}
\newcommand{\be}{\begin{equation}}
	\newcommand{\ee}{\end{equation}}
\newcommand{\bea}{\begin{eqnarray}}
	\newcommand{\eea}{\end{eqnarray}}

\newcommand{\MEG}{\mu \rightarrow e \gamma}

\title{Light thermal dark matter via type-I seesaw portal}
\author[a]{Debasish Borah,}
\emailAdd{dborah@iitg.ac.in}
\author[a,b]{Pritam Das,}
\emailAdd{prtmdas9@gmail.com}
\author[c]{Satyabrata Mahapatra,}
\emailAdd{satyabrata@g.skku.edu}
\author[d]{Narendra Sahu,}
\emailAdd{nsahu@phy.iith.ac.in}

\affiliation[a]{Department of Physics, Indian Institute of Technology, Guwahati, Assam 781039, India.}
\affiliation[b]{Department of Physics, Salbari College, Baksa, Assam 781318, India.}
\affiliation[c]{Department of Physics and Institute of Basic Science, Sungkyunkwan University, Suwon 16419, Korea.}
\affiliation[d]{Department of Physics, Indian Institute of Technology Hyderabad, Kandi, Sangareddy 502285, Telangana, India.}

\abstract{We propose a minimal scenario for light thermal dark matter (DM) in sub-GeV to GeV range by incorporating a scalar singlet DM in a type-I seesaw scenario extended by an additional Higgs doublet $\phi_2$. The latter permits efficient annihilation of light scalar DM into leptonic final states including right-handed neutrinos (RHN).
We keep the charged fermion as well as neutral fermion final states in forbidden regime to avoid bounds from indirect search as well as cosmic microwave background (CMB) data. After studying the purely forbidden DM scenario with neutral and charged fermion final states separately, we discuss the interplay of forbidden and non-forbidden channels in generating light thermal DM relic by considering neutral fermions in non-forbidden mode due to relatively weaker constraints. The model can also explain the anomalous magnetic moment of muon, W-mass anomaly and saturate experimental bounds on charged lepton flavour violation and DM direct detection while offering tantalising detection prospects of RHN, 
 the mass of which is kept approximately in the same range as DM.}
\begin{document} 
	\maketitle
		\flushbottom
	
	\setcounter{footnote}{0}
	\renewcommand*{\thefootnote}{\arabic{footnote}}
	%%%%%%%%%%%%%%%%%%%%%%%%%%%%%%%%%%%%%%%%%%%%%%%%%%%%%%%%%%%%%%%%%%%%%%%%%%%%%%%%%%%%%%%%%%%
	\section{Introduction}
The matter component in the present Universe is dominated by a non-luminous, non-baryonic form of matter, popularly known as dark matter (DM). This has been supported by various astrophysical observations at different scales \cite{Zwicky:1933gu, Rubin:1970zza, Clowe:2006eq} together with cosmological 
experiments like PLANCK, WMAP predicting around 26.8\% of the present Universe to be made up of DM \cite{Zyla:2020zbs, Planck:2018vyg}. In terms of density 
parameter $\Omega_{\rm DM}$ and reduced Hubble parameter $h = \text{Hubble Parameter}/(100 \;\text{km}
~\text{s}^{-1} \text{Mpc}^{-1})$, the observed DM abundance in the present epoch at 68\% CL
is \cite{Planck:2018vyg}
\begin{equation}
\Omega_{\text{DM}} h^2 = 0.120\pm 0.001 
\label{dm_relic}\,.
\end{equation}
Given DM has a particle origin, none of the standard model (SM) particles can satisfy the required criteria of a particle DM. This has led to several beyond the standard model (BSM) proposals for DM out of which the weakly interacting massive particle (WIMP) has been the most popular one. In the WIMP paradigm, a DM particle having mass and interactions similar to those around the electroweak scale gives rise to the observed relic after thermal freeze-out, a remarkable coincidence often referred to as the {\it WIMP Miracle} \cite{Kolb:1990vq}. A recent review of such models can be found in \cite{Arcadi:2017kky}. Typically, the interactions leading to thermal freeze-out of WIMP also give rise to sizeable DM-nucleon scattering which has been searched for at several direct detection experiments. However, no such scattering has been observed yet leading to stringent constraints on WIMP DM parameter space \cite{LUX-ZEPLIN:2022qhg}.

In view of this, light thermal DM with mass $(M_{\rm DM} \lesssim \mathcal{O}(10 \, \rm GeV))$ has received lots of attention in recent times, particularly due to weaker constraints from direct detection experiments like LZ \cite{LUX-ZEPLIN:2022qhg}. However, it is difficult to achieve the {\it WIMP Miracle} in such a low mass regime typically due to insufficient annihilation rate of DM leading to thermal overproduction. For fermionic DM, the criteria of thermal DM not overclosing the Universe leads to a lower bound on its mass, around a few GeV \cite{Lee:1977ua, Kolb:1985nn}. Related discussions and exceptions for scalar DM can be found in \cite{Boehm:2003hm}. In the presence of light mediators between DM and SM sectors, however, one can achieve the correct relic abundance as pointed out in several works \cite{Pospelov:2007mp, DAgnolo:2015ujb, Berlin:2017ftj, DAgnolo:2020mpt, Herms:2022nhd, Jaramillo:2022mos}\footnote{See also Refs~\cite{Spergel:1999mh,Tulin:2017ara,Borah:2023sal, Borah:2022ask, Borah:2021yek, Borah:2021pet, Borah:2021rbx, Borah:2021qmi,Dutta:2022knf} where a large annihilation cross-section is achieved due to a light mediator introduced to explain DM self-interactions.}. However, such light DM with a large annihilation rate to SM often faces tight constraints from cosmic microwave background (CMB) observations \cite{Madhavacheril:2013cna, Slatyer:2015jla, Planck:2018vyg}. Such constraints can be evaded if DM is kept in the kinematically forbidden regime \cite{DAgnolo:2015ujb, DAgnolo:2020mpt, Griest:1990kh}.

Motivated by this, we consider a simple realisation of light thermal DM in a type-I seesaw framework extended by a second Higgs doublet. DM annihilates dominantly via the light neutral component of this additional Higgs doublet into light neutrinos and the heavy right-handed neutrino (RHN) of mass ranging from MeV to GeV. We have considered the mass hierarchy of the RHNs as $M_{N_1}<M_{N_2}<<M_{N_3}$.
Such RHNs can take part in type-I seesaw mechanism \cite{Minkowski:1977sc, GellMann:1980vs, Mohapatra:1979ia, Schechter:1980gr}, leading to the generation of light neutrino masses and mixing, another observed phenomenon SM fails to address. Unlike in \cite{Herms:2022nhd} where forbidden DM mass was close to muon or tau lepton masses (also to other SM particles studied in \cite{DAgnolo:2020mpt}), here we can have a wide range of DM masses due to the freedom in choosing lightest RHN mass. We consider the RHN-SM neutrino final state in kinematically forbidden mode while calculating the relic of DM by solving the relevant Boltzmann equations numerically. While neutral fermion final states are less constrained from indirect detection or CMB bounds, the subsequent decay of RHN into charged leptons and photons can be constrained from present and future observations \cite{Escudero:2016tzx, Escudero:2016ksa, Coito:2022kif, Morrison:2022zwk}. We first study the scenario where the second Higgs doublet is neutrinophilic such that DM primarily annihilates into RHN-SM neutrinos. We then consider the second Higgs doublet to be leptophilic such that charged fermion final states like muons also become viable. While DM relic is primarily governed by the forbidden channels, the scenario can still be probed at future direct detection experiments searching for DM-electron or DM-nucleon scatterings. While we discuss the possibility of light scalar singlet DM in this work, it is also possible to study fermion singlet DM by introducing dimension five operators leading to DM annihilation via the neutral component of the second Higgs doublet.

Due to the existence of a light scalar component of the second Higgs doublet as well as light RHN, the model remains verifiable via heavy neutral lepton (HNL) search experiments, charged lepton flavour violation in addition to collider aspects of the second Higgs doublet. The light CP-even scalar component of the second Higgs also gives rise to a positive contribution to muon $(g-2)$ while the negative contribution from the one-loop diagram mediated by charged scalar and CP-odd scalar is tuned to be sub-dominant while being consistent with the neutrino mass and lepton flavour violation (LFV) constraints. By properly choosing the masses of the charged and CP-odd scalar while CP-even scalar is still light as per the requirement of achieving correct relic density of DM, it is also possible to explain the CDF-II W mass anomaly~\cite{CDF:2022hxs} by the self-energy correction of W-boson mass with the new doublet scalars in the loop.

This paper is organised as follows. In section \ref{sec1} we briefly describe our model followed by a discussion of neutrino mass in section \ref{sec2}. In section \ref{sec:g2}, we discuss the details of muon $(g-2)$ and charged lepton flavour violation followed by the discussion of CDF-II W-mass anomaly in section \ref{sec:wmass}. In section \ref{sec4}, we present the results related to light thermal DM followed by detection prospects of heavy neutral leptons in section \ref{sec6}. We finally conclude in section \ref{sec7}.

 \section{The Model}
 \label{sec1}
\begin{figure}[h]
    \centering
    \includegraphics[scale=0.5]{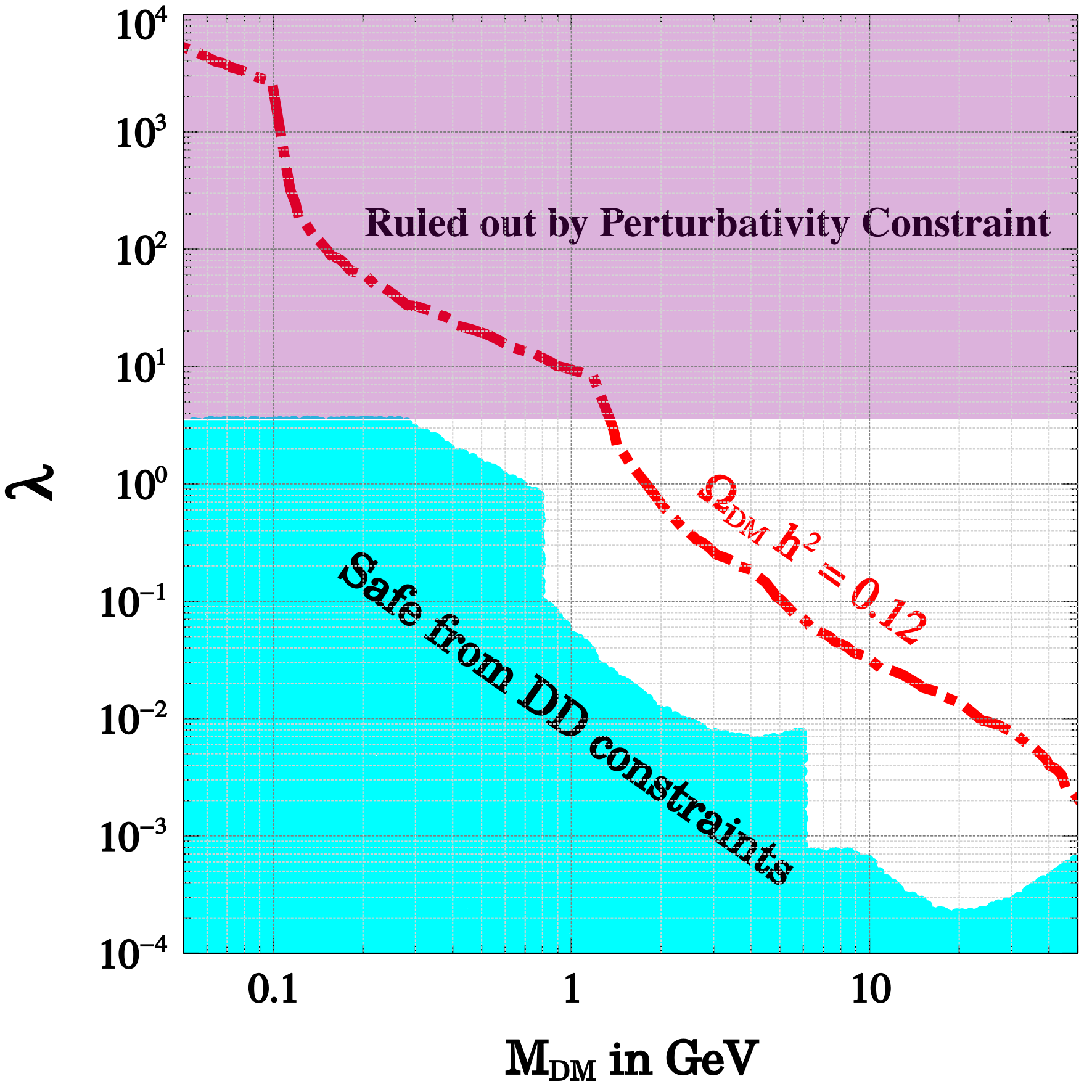}
    \caption{Parameter space for singlet scalar DM extension of the SM in sub-GeV to GeV scale DM mass window.}
    \label{fig:singletscalar}
\end{figure}

As we are going to discuss the possibility of light scalar singlet DM in a type-I seesaw scenario extended by a second Higgs doublet, we briefly comment upon the status of scalar singlet DM extension of the SM. In Fig.~\ref{fig:singletscalar}, we showcase the parameter space for a singlet scalar DM scenario in the plane of the singlet scalar DM coupling with SM Higgs ($\lambda S^2 (\phi_1^\dagger \phi_1)$) and DM mass. The red dot-dashed line shows the contour of correct relic density considering the annihilation cross-section of $S$ into SM fermions mediated via SM Higgs. Clearly it is difficult to achieve correct relic density below a few GeV of DM mass while being consistent with the perturbativity constraint on the coupling $\lambda$ which is depicted by the purple shaded region. We also show the parameter space consistent with the most stringent constraint on DM-nucleon scattering from CRESST-III~\cite{CRESST:2019jnq}, DS-50~\cite{DarkSide-50:2022qzh}, XENON-nT~\cite{XENON:2023cxc} and LZ~\cite{LUX-ZEPLIN:2022qhg} depending on the DM mass with the cyan shaded region. It is evident from Fig.~\ref{fig:singletscalar} that for DM mass below $50$ GeV, there is no common parameter space that satisfies correct relic density and direct detection constraints. The coupling required to achieve sufficient annihilation cross-section so as to get the correct relic density are already ruled out by direct search experiments.

 We consider a type-I seesaw model extended by a Higgs doublet $(\phi_2)$ and a real singlet scalar $(S)$ with the latter being odd under an unbroken $Z_2$ symmetry. The singlet scalar, being stable due to $Z_2$ symmetry, acts like a DM candidate in the model. Thus, the scalar sector consists of the SM Higgs doublet $\phi_1$, the second Higgs doublet $\phi_2$ and the $Z_2$-odd scalar singlet $S$. In order to keep $Z_2$ symmetry unbroken, $S$ does not acquire any vacuum expectation value (VEV). We consider the alignment limit of the two Higgs doublets, where only one neutral Higgs (SM-like) acquires a non-zero VEV $(v)$ such that they can be parameterised as
 \begin{equation}
     \phi_{1}=\begin{pmatrix}G^+\\\frac{1}{\sqrt{2}}(v+h_1+iG^0)\end{pmatrix}\quad 
     \phi_{2}=\begin{pmatrix}H^+\\\frac{1}{\sqrt{2}}(H_2+iA)\end{pmatrix}
 \end{equation}
 The scalar potential of the model can be written as follows,
	\begin{eqnarray}
		V=&&\nonumber-\mu_{\phi_1}^2 \phi_1^\dagger \phi_1 + \mu_{\phi_2}^2 \phi_2^\dagger \phi_2 + \frac{1}{2}\mu_{ S}^2 S^2 +\frac{1}{2}\lambda_1 (\phi_1^\dagger \phi_1)^2 +\frac{1}{2} \lambda_2 (\phi_2^\dagger\phi_2)^2 +\frac{1}{4!} \lambda_3 S^4 \\&&+ 
		\lambda_3 (\phi_1^\dagger \phi_1) (\phi_2^\dagger\phi_2) + \lambda_4 (\phi_1^\dagger \phi_2) (\phi_2^\dagger \phi_1) + \frac{1}{2}\lambda_5 (\phi_1^\dagger \phi_2) (\phi_1^\dagger \phi_2)\\&&+\frac{1}{2} \lambda_{S1} (\phi_1^\dagger \phi_1) S^2 +\frac{1}{2}\lambda_{S2} (\phi_2^\dagger\phi_2) S^2  + 
		\lambda_{HS} (\phi_2^\dagger \phi_1+\phi_1^\dagger \phi_2) S^2\nonumber.
	\end{eqnarray}
	We work on the Higgs basis, where one of the neutral Higgs mass eigenstates is aligned with the direction of the  VEV of the scalar field. From previous studies  \cite{Bernon:2015qea, Bernon:2015wef, Babu:2019mfe}, it is clear that for two Higgs doublet cases, the alignment limit is independent of the choice of basis and we have considered it to be exhibited in the Higgs basis itself. The scalar doublet $\phi_1$ has tree-level couplings to the SM particles. Therefore, if one of the CP-even neutral Higgs mass eigenstates is SM-like, then it must be approximately aligned with the real part of the neutral field $h_1$.
Hence, in the alignment limit\footnote{We have considered zero mixing between these two doublets (in this case, the $Z_2$ basis and Higgs basis coincide and the quartic couplings for $(\phi^\dagger_i\phi_i)\phi^\dagger_j\phi_i$) terms will be zero \cite{Bernon:2015wef,Haber:2015pua}. The $Z_2$-basis corresponds to an approximate $Z_2$ symmetry obeyed by the scalar potential only with $\phi_2$ being $Z_2$-odd.} \cite{Bernon:2015qea} the SM Higgs ($h_1\sim h$) decouples from the new CP-even Higgs ($H_2$) and the mass spectrum for the physical scalars, can be obtained as follows,
	\begin{eqnarray}
		M_h^2=&&\lambda_1 v^2;\quad 	M_{A}^2=m_{H_2}^2-\lambda_5 v^2;\quad             M_S^2=\mu^2_S+\lambda_{S1}v^2;\\
		M_{H_2}^2=&&\mu_{\phi_2}^2+\frac{v^2}{2}(\lambda_3+\lambda_4+\lambda_5);	
		\quad M_{H^\pm}^2=M_{H_2}^2-\frac{v^2}{2}(\lambda_4+\lambda_5).
	\end{eqnarray}

From the above equations, it is evident that by considering $\lambda_4$ and $\lambda_5$ of order $\mathcal{O}(1)$, it is possible to create a large mass difference between $M_{H_2}$ and $M_{H^{\pm},A}$ (in order to satisfy electroweak precision bounds \cite{Lundstrom:2008ai}) and we exploit this fact to realise light forbidden DM while concurrently achieving required positive and negative contributions to anomalous magnetic moment of muon and electron respectively which is discussed in subsequent sections.

With the inclusion of three right-handed neutrinos (all are $Z_2$-even), the new terms in the Yukawa Lagrangian for this model can be expressed as,
\begin{eqnarray}\label{lag1}
    -\mathcal{L}\supset y^{\alpha k}_{1}\Bar{L_\alpha}\tilde{\phi_1} N_k+y^{\alpha k}_{2}\Bar{L_\alpha}\tilde{\phi_2} N_k+Y^{\alpha}_{2}\Bar{L_\alpha}{\phi_2} l_{R_{\alpha}} + {\rm h.c.},
\end{eqnarray}
where, $\alpha=e,\mu,\tau$ and $k=1,2,3$.

The first term in the above Lagrangian gives rise to the neutrino mass generation through the type-I seesaw mechanism whereas the second term is relevant for the forbidden DM realisation with type-I seesaw portal. It is worth noting here that, this term also leads to a one-loop contribution to neutrino mass similar to the scotogenic model \cite{Tao:1996vb, Ma:2006fn}. We also consider the charged lepton Yukawa coupling with the second Higgs doublet $\phi_2$ to be of diagonal type to avoid tree-level flavour changing neutral current. We are also assuming the RHN mass matrix to be diagonal for simplicity. 

Our study is divided into two categories depending upon whether the second Higgs doublet $\phi_2$ is neutrinophilic or leptophilic. For neutrinophilic scalar doublet $\phi_2$, only the first two terms of the interaction Lagrangian in Eq. \eqref{lag1} exist. In this case, the light neutrino masses are obtained by the combined contribution from the tree-level as well as the one-loop level. In the dark matter phenomenology, only the $SS\rightarrow N_1 \bar{\nu}_\alpha$ annihilation channels mediated by light $H_2$ dominate. However, if the $\phi_2$ is assumed to be leptophilic then it can couple to charged leptons in addition to the neutrinos governed by the third term in Eq.~\eqref{lag1}. This facilitates the model to explain the muon $(g-2)$ anomaly and also enhances the detection prospects at the LFV experiments like MEG, Mu3e etc. 

In the leptophilic scenario, DM relic density is dominantly decided by the $SS\rightarrow \ell\bar{\ell}$ channels in the forbidden regime. However, in the non-forbidden regime, $SS\rightarrow N_1 \bar{\nu}_\alpha$ will dominate, as we discuss in upcoming sections. In both scenarios, we get a strong correlation between the neutrino mass, DM phenomenology and the flavour observables.

\section{Neutrino Mass}
\label{sec2}
In this setup, the mass of the active neutrino is generated through both tree-level and one-loop processes. At the tree-level, the active neutrino mass is generated through the type-I seesaw mechanism following the breaking of electroweak symmetry. On the other hand, at the one-loop level, the mass arises from the involvement of $N_i$ and $\phi_2$ particles within the loop, resembling the scotogenic origin \cite{Tao:1996vb, Ma:2006fn}. It is worth noting that, in this scenario, neither of the particles within the loop are considered potential dark matter candidates due to the absence of any exact symmetry ensuring their stability. The relevant Lagrangian for neutrino mass 
is given by

\begin{eqnarray}
    \mathcal{L} \supset -y^{\alpha k}_{1}\Bar{L_\alpha}\tilde{\phi_1} N_k-y^{\alpha k}_{2}\Bar{L_\alpha}\tilde{\phi_2} N_k- \frac{1}{2} (\bar{N^c_k}M_{N_k} N_k)+ {\rm h.c.}
\end{eqnarray}
The type-I seesaw contribution to neutrino mass is given as,
\begin{eqnarray}
    (m_\nu^{\alpha \beta})_{\rm tree}=-M_DM_N^{-1}M_D^T\equiv -\frac{1}{2}y^{\alpha k}_1 M_{N_k}^{-1} y^{k \beta}_1 v^2;\quad\quad %\theta=M_DM_N^{-1}
    \label{as1}
\end{eqnarray}
with $M_D$ being the Dirac mass term, which can be parameterised as $M_D=y_1 \frac{v}{\sqrt{2}} $. To ensure the connection between light neutrino oscillation parameters and the active-sterile mixing angle $\theta$ originating from type-I seesaw, we have adopted the Casas-Ibarra(CI) parameterisation in type-I seesaw \cite{Drewes:2019mhg, Casas:2001sr}
\begin{eqnarray}
    \theta_{\alpha k}=\left(iU_{\rm PMNS}\sqrt{m_{\nu}^{\rm Diag}}\mathcal{R} \sqrt{M_N^{-1}}\right)_{\alpha k}
    \label{ciparam}
\end{eqnarray}
where, $U_{\rm PMNS}$ is the unitary Pontecorvo-Maki-Nakagawa-Sakata (PMNS) leptonic mixing matrix\footnote{We assume the individual seesaw mass matrices to be diagonalized by the leptonic mixing matrix for simplicity. The charged lepton mass matrix is considered to be diagonal.}, $m_{\nu}^{\rm Diag}$ and $M_{N}$ are the $3\times3$ diagonal light neutrino and heavy neutrino mass matrices, respectively. Here, $\mathcal{R}$ is an arbitrary complex orthogonal matrix, with $\mathcal{R}\mathcal{R}^T=1$.

We also get a one-loop contribution to neutrino mass with $\phi_2$ and $N$ in the loop which is given by \cite{Ma:2006fn, Merle:2015ica}:
\begin{eqnarray}
    (m_\nu^{\alpha \beta})_{\rm loop}=\sum\frac{y_2^{\alpha k}y_2^{k \beta}M_{N_k}}{32\pi^2}\Big[f_k(M_{H_2}^2)-f_k(M_{A}^2)\Big],
\end{eqnarray}
where, $M_{N_k}$ is the mass eigenvalue of the RHN mass eigenstate $N_k$ and the loop function is defined as, $f_k(M^2_x)=\frac{M_x^2}{M_x^2-M_{N_k}^2}{\rm ln}\frac{M_x^2}{M_{N_k}^2}$. 
Therefore the total neutrino mass will be the sum of both tree-level and loop-level contributions, $i.e.$,
\begin{eqnarray}
   m_\nu^{\alpha \beta}= (m_\nu^{\alpha \beta})_{\rm tree}+ (m_\nu^{\alpha \beta})_{\rm loop}= \sum \Big[\frac{-y_1^{\alpha k} y_1^{k \beta} v^2}{2M_{N_k}}+ \frac{y_2^{\alpha k}y_2^{k\beta}M_{N_k}}{32\pi^2}\big[f_k(M^2_{H_2})-f_k(M_{A}^2)\big]\Big].
\end{eqnarray}

In our analysis, we assume equal weightage of the two contributions to the neutrino mass, $i.e.$, both the tree-level and loop-level contributions collectively account for approximately $50\%$ of the total light neutrino mass. We adopt a bottom-up approach in determining the neutrino mass to keep the whole analysis consistent. Initially, we use the most recent best-fit values for the neutrino parameters as per \cite{Esteban:2020cvm} to formulate the total light neutrino mass matrix. Subsequently, employing the CI parameterisation within the framework of the type-I seesaw (with a consideration of $50\%$ contribution from tree-level neutrino masses), we derive the Yukawa couplings ($y_2$) governing the one-loop contribution. A more general parameterisation of the individual seesaw contributions will not drastically change the generic conclusions arrived at in our work.

\section{Muon $(g-2)$ and LFV}
\label{sec:g2}
\begin{figure}[h]
    \centering
    \includegraphics[scale=0.25]{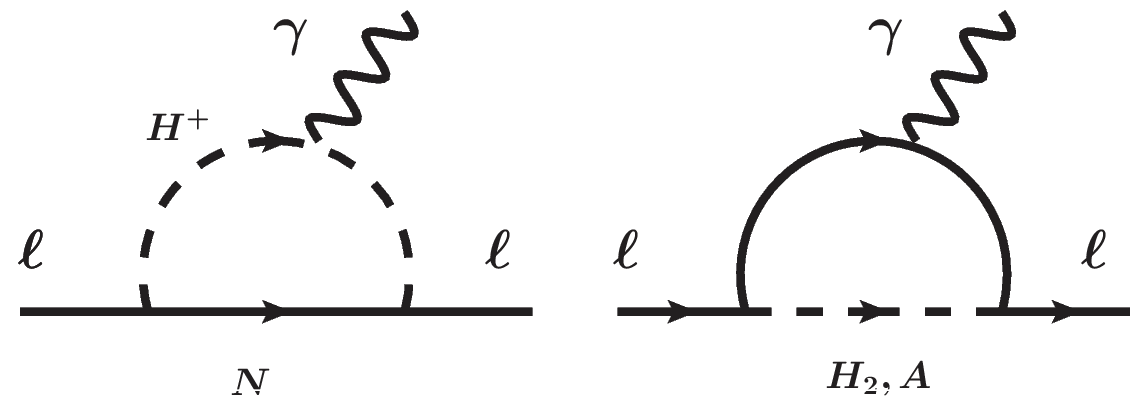}
    \caption{One-loop diagrams contributing to the magnetic moment of leptons.}
    \label{lfv}
\end{figure}

In this model, the presence of the second Higgs doublet offers the possibility to account for the anomalous magnetic moment of the muon $a_\mu$ = $(g - 2)_\mu/2$ by virtue of the loop diagrams shown in Fig. \ref{lfv}. The Muon g-2 collaboration at the Fermilab has recently reported\cite{Muong-2:2023cdq} $\Delta a_\mu = a^{\rm exp}_\mu - a^{\rm SM}_\mu = 249 (48)\times 10^{-11}$, a discrepancy of $5.1\sigma$ CL. While recent lattice results \cite{Borsanyi:2020mff} can alleviate this discrepancy to some extent, there still remains scope for BSM physics to play the leading role in explaining this discrepancy \cite{Jegerlehner:2009ry, Lindner:2016bgg, Athron:2021iuf}.

The new contributions to $(g-2)_\mu$ with the neutral scalars in the loop is given by\cite{Lindner:2016bgg}
\begin{subequations}
\begin{eqnarray}
  && \Delta a_\mu(H_2) = \frac{1}{8\pi^2} \frac{m_\mu^2}{M_{H_2}^2}\int_0^1 \mathrm{d}x\,  \frac{\left(Y^\mu_{2}\right)^2 x^2( 2-x)}{(1-x)(1-x \left(\frac{m_\mu}{M_{H_2}}\right)^2)+x\left(\frac{m_\mu}{M_{H_2}}\right)^2}   \\
 &&  \Delta a_\mu(A) = -\frac{1}{8\pi^2} \frac{m_\mu^2}{M_{A}^2}\int_0^1 \mathrm{d}x\,  \frac{\left(Y^\mu_{2}\right)^2 (x^3)}{(1-x)(1-x \left(\frac{m_\mu}{M_{A}}\right)^2)+x\left(\frac{m_\mu}{M_{A}}\right)^2} 
\end{eqnarray}
\end{subequations}
and the contribution from the charged scalar and RHN loop is given by

  \begin{equation}
    \Delta a_\mu\left(H^+\right) = -\frac{1}{8\pi^2} \frac{m_\mu^2}{M_{H^+}^2}\int_0^1 \mathrm{d}x\, \sum_k \frac{\left|y^{\mu k}_{2}\right|^2 2x^2(1-x)}{\left(\frac{M^2_{N_k}}{M^2_{H^+}}\right)(1-x)\left(1-\left(\frac{m^2_\mu}{M^2_{N_k}}\right)x\right)+x}.\label{eq:Delta_a_Singly_Scalar}
  \end{equation}

Here, we observe three novel contributions from BSM physics affecting the muon anomalous magnetic moment, stemming from one-loop diagrams involving $H_2$, $A$, and $H^{+}$ particles in the loop. The contribution originating from the $H_2$ loop yields a positive contribution, whereas those involving $A$ and $H^{+}$ yield negative impacts. Given the muon anomalous magnetic moment being reportedly positive, precise tuning of the masses of $H_2$, $A$, and $H^+$ and their corresponding couplings is required to yield an overall positive  $\Delta a_\mu$. Clearly, to achieve the accurate $\Delta a_\mu$, it is imperative for the contribution from $H_2$ to surpass that of $A$ and $H^{+}$.

The same particles in the loop can also contribute to LFV decays like $\mu \rightarrow e \gamma$. Since we have assumed charged lepton coupling with $\phi_2$ to be diagonal, we get the significant new physics contributions only from the charged scalar and RHN loop. The branching ratio for the $\mu \rightarrow e \gamma$ process mediated via the charged scalar can be estimated as \cite{Lindner:2016bgg}
\begin{eqnarray}
 \mathrm{BR}(\MEG)\approx \frac{3(4\pi)^3 \alpha_\mathrm{em}}{4 G_F^2}\left( |A_{e\mu}^M|^2 + |A_{e\mu}^E|^2 \right) ,  
 \label{brmeg}
\end{eqnarray}
where the form factors are defined as follows
\begin{subequations}\label{eq:BR2}
  \begin{align}
    A_{e\mu}^M &= \frac{-1}{(4\pi)^2}\sum_k \left[ {y_{2}^{ke}}^*y_{2}^{k\mu} (G^{+}+ G^{-})\right]\,,\\ 
    A_{e\mu}^E &= \frac{i}{(4\pi)^2}\sum_k \left[{y_{2}^{ke}}^*y_{2}^{k\mu} (G^{+} + G^{-})\right]\,,
  \end{align}
\end{subequations}
with the loop function being
\begin{equation}
  G^{\pm} \simeq \frac{1}{M_{H^+}^2}\int_0^1 \mathrm{d}x \int_0^1 \mathrm{d}y \, x(1-x) \frac{xy \pm \left(\frac{M_{N_k}}{m_\mu}\right)}
  {\left(\frac{M^2_{N_k}}{M^2_{H^+}}\right)(1-x)\left(1-\left(\frac{m^2_\mu}{M^2_{N_k}}\right)xy\right)+x}.
\end{equation}

\begin{figure}[h]
    \centering
    \includegraphics[scale=0.46]{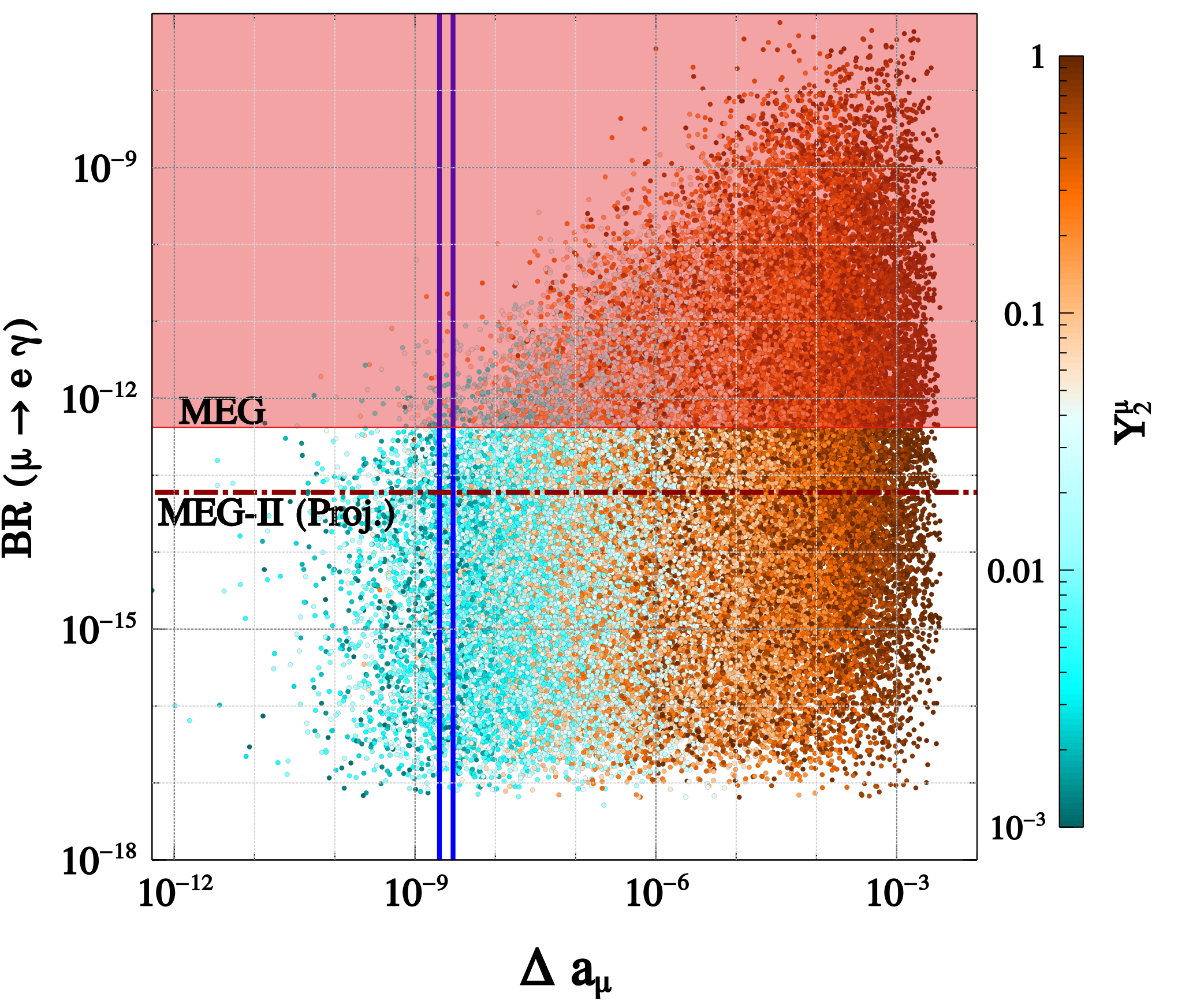}\hfil
 \includegraphics[scale=0.45]{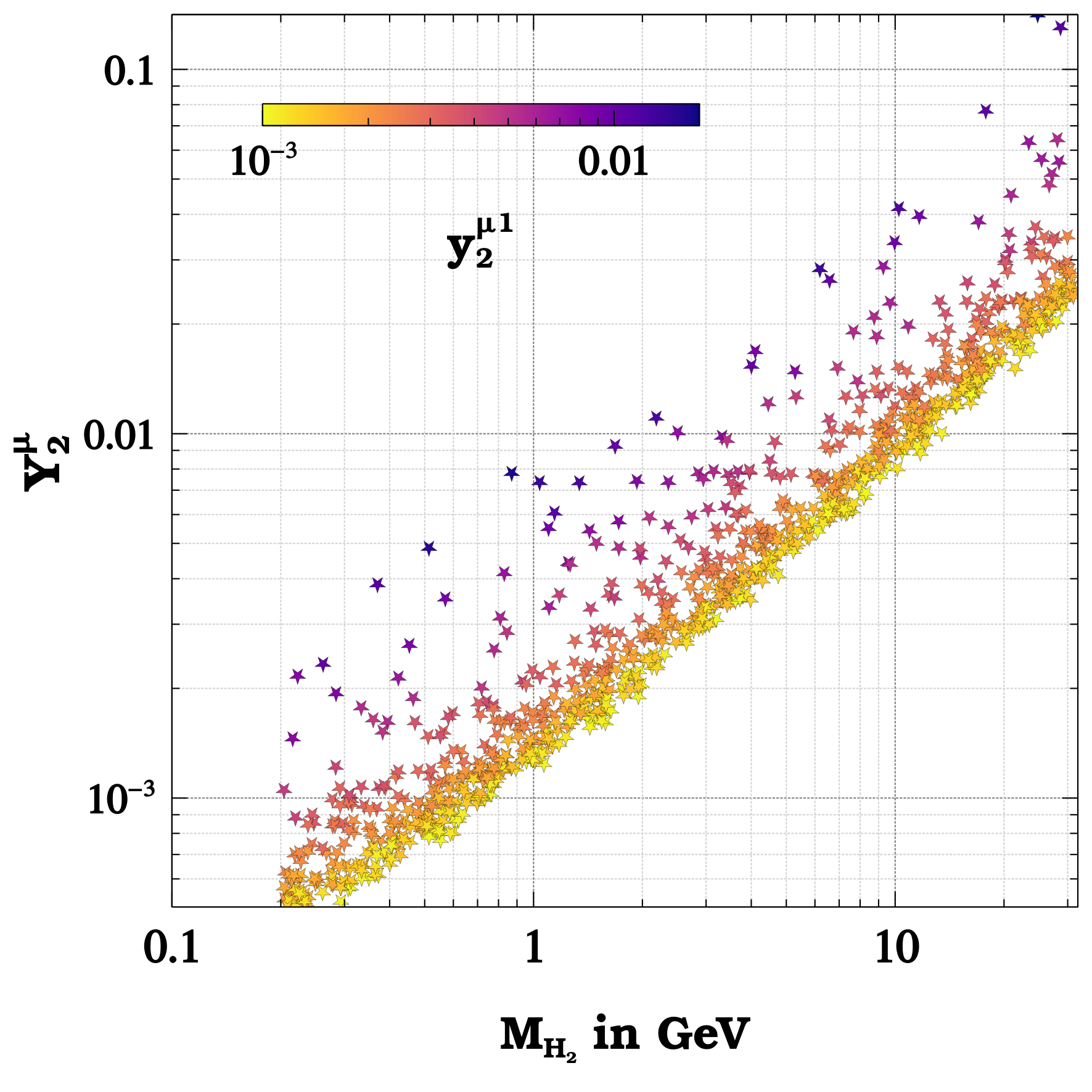}
\caption{{ Left: Parameter scan points in the plane of ${\rm BR}(\mu\to e\gamma)$ and $\Delta a_\mu$.} Right: Parameter space giving rise to correct $\Delta a _{\mu}$ consistent with the most recent constraint on BR$(\mu\to e\gamma)$ from MEG-II~\cite{Cattaneo:2023iua}}
    \label{g2check}
\end{figure}

{
 \begin{table}[h]
     \centering
   {\begin{tabular}{|c|c|}
          \hline
          \hline
  Parameters&  Range  \\
         \hline
         \hline
         $M_{\rm DM}$(GeV)& Neutrinophilic:$[0.5,5]$\\ & Leptophilic: $[0.05,0.105]$  \\
         $M_{H_2}$(GeV)& $[0.2,30]$\\
         $M_{A}$(GeV)& $[100,300]$\\
         $M_{H^+}$(GeV)& $[100,300]$\\
         $\Delta$&$[10^{-4},0.1]$\\
         $Y_2^\mu$&$[10^{-4},1]$\\
        
         \hline
     \end{tabular}}
     \caption{The range in which the free parameters are randomly varied for the numerical analysis.}
     \label{tab:param_range}
 \end{table}
}

In Eq. \eqref{brmeg}, $G_F$ is the Fermi constant and $\alpha_{\rm em}$ is the fine-structure constant. From the experimental point of view, the current upper limits on the $\mu \rightarrow e\gamma$ branching ratio from MEG-2016 result is BR$(\mu\rightarrow e\gamma)<4.2\times10^{-13}$ \cite{MEG:2016leq}, with future sensitivity being BR$(\mu\rightarrow e\gamma)<6\times10^{-14}$ \cite{MEGII:2018kmf}. {We perform a random scan of the parameters to find out the parameter space that can give rise to the required $\Delta a_\mu$ while still being consistent with the LFV constraints. The range in which we vary these free parameters randomly are mentioned in Table~\ref{tab:param_range}}.
{ In the left panel of Fig. \ref{g2check}, we showcase the results of our parameter scan in the plane of BR$(\mu\rightarrow e\gamma)$ and $\Delta a_\mu$.  It is interesting to see that a part of the parameter space that can explain the muon $g-2$ anomaly can be probed by the future sensitivity of LFV experiments like MEG-II \cite{MEGII:2018kmf}.  } The right panel of Fig. \ref{g2check} shows the parameter space in terms of Yukawa couplings of the muon with $\phi_2$ and light neutral CP-even scalar mass which can explain the anomalous muon $(g-2)$ while being consistent with the MEG upper limits on $\MEG $. The Yukawa coupling of the muon with the charged scalar and RHN is shown in the colour code. 
Since $H_2$ mass is smaller as compared to $A$ and $H^+$, it is possible to achieve correct $\Delta a_\mu$ through the one-loop diagram involving $H_2$ while the negative contribution from $H^+$ and $A$ loop remains suppressed. 
{Here it should be mentioned that, while varying the scalar masses $M_A$ and $M_{H^+}$ we ensure the perturbativity of the scalar couplings.  The masses of RHNs are varied such that $M_{N_2} > 2 M_{\rm DM}$. ({\it i.e.} $M_{N_2} = 2 M_{\rm DM} (1+\Delta)$ with $\Delta$ varied as mentioned in Table~\ref{tab:param_range}), $M_{N_1} < 2 M_{\rm DM}$ and $M_{N_3}=M_{N_2}+100$ GeV. The reason behind such a choice will be clear when we discuss the forbidden DM scenario in subsequent sections.}

\section{W-Mass Anomaly}
\label{sec:wmass}

In our scenario, a positive contribution to the $\rho$ parameter can come from the self-energy correction of the W-boson with the new doublet scalar. This additional contribution to self-energy correction $\Delta \rho$ and hence the $T$- parameter ($= {\Delta \rho}/{\alpha_{\rm em}}$) is given by \cite{Babu:2022pdn,Borah:2023hqw}: 
	\begin{eqnarray} 
	T  =\scriptstyle \dfrac{{\Theta}(M_{H^+}^2,M_{H_2}^2) + \Theta(M_{H^+}^2,M_{A}^2) 
		-\Theta(M_{H_2}^2,M_{A}^2)}{16\pi^2 \alpha_{\rm em}(M_Z) v^2 }\,, \label{eq:T}\nonumber\\
	\end{eqnarray}
	with the loop function $\Theta$ given by:
	\begin{equation} \label{Fdef}
	\Theta(x,y) \  \equiv \  \frac{1}{2}(x+y) -\frac{xy}{x-y}\ln\left(\frac{x}{y}\right)\,.
	\end{equation}

\begin{figure}[h]
    \centering
    \includegraphics[scale=0.5]{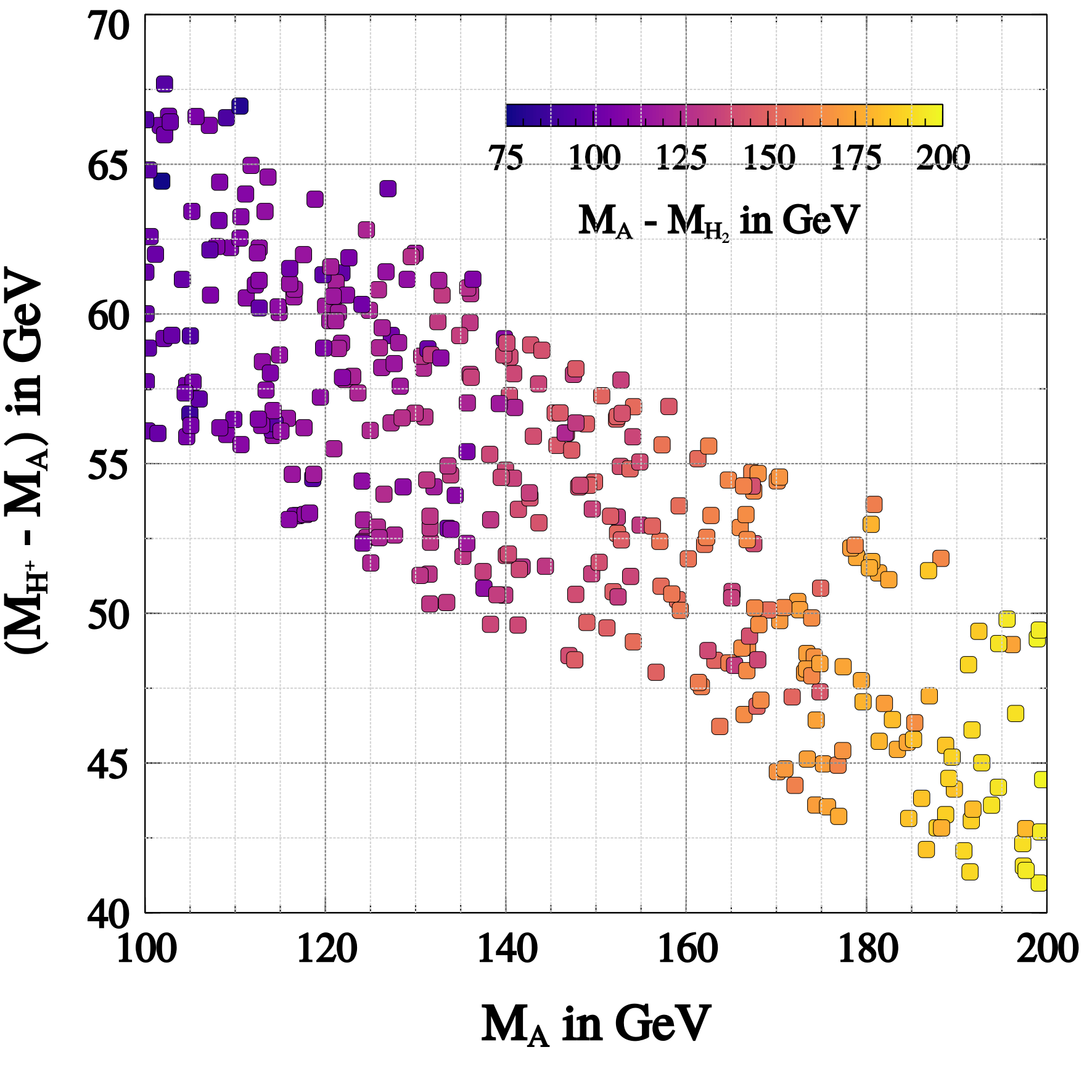}
    \caption{Parameter space in the plane of doublet scalar masses that can explain CDF-II W-mass anomaly while being consistent with the muon $(g-2)$ and LFV constraints.}
    \label{fig:mwcdf}
\end{figure}
 
	In addition to the $T$-parameter contribution, the $S$-parameter can also modify the $W$-boson mass slightly. The $S$ parameter is given as,
	\begin{eqnarray}
	S=\frac{1}{12\pi}\log\bigg[\frac{M_{H_2}^2+M_{A}^2}{2M_{H^+}^2}\bigg].
	\end{eqnarray}
	The modified $W$-boson mass considering both these contributions is given by \cite{Grimus:2008nb,Borah:2022obi}
	\begin{align} 
	& M_W  \simeq M^{SM}_{W}\left[1- \dfrac{\alpha_{\rm em}(M_Z)(S-2~\cos^2{\theta_W}~ T)}{4(\cos^2{\theta_W}-\sin^2\theta_W)}\right]. \label{eq:MW-S}
	\end{align}

	We observe that the alteration in the W-boson mass caused by the $S$ parameter is generally negligible, with the primary correction arising dominantly from the $T$ parameter. In Fig.~\ref{fig:mwcdf}, we showcase the parameter space in the plane of $M_A$ and $(M_{H^+}-M_A)$ with $(M_A-M_{H_2})$ depicted in the color code. Note that all these points also satisfy the correct muon $(g-2)$ as well as is consistent with the constraints from LFV experiments as discussed in the section~\ref{sec:g2}. Clearly it is possible to explain the CDF-II W-mass anomaly~\cite{CDF:2022hxs} with $M_{H_2}\in[0.2,30]$  GeV and $M_{A}\in [100,200]$ GeV while $M_{H^{+}} \in M_{A}+[40,70]$ GeV.

\section{Light thermal dark matter}
\label{sec4}

While the WIMP paradigm is straightforward \cite{Kolb:1990vq}, in kinematically forbidden DM scenarios \cite{DAgnolo:2015ujb, DAgnolo:2020mpt, Griest:1990kh}, the mass of the final-state particles to which DM annihilates into in the early Universe exceeds that of the initial state or DM particles. This is made feasible by fixing a small mass difference between the initial and final-state particles, thereby introducing the essential Boltzmann suppression factor required to achieve the appropriate dark matter relic abundance. Such forbidden DM freezes out earlier compared to standard thermal WIMP DM due to Boltzmann suppression associated with heavier final states at lower temperatures. One can still satisfy the correct relic by resonantly enhancing the annihilation channels in the forbidden regime.
We consider this possibility in our model where light-forbidden DM $S$, annihilates via light mediator (a neutral component of second scalar doublet $\phi_2$) into different final states depending upon neutrinophilic or leptophilic nature of $\phi_2$. We discuss these two sub-cases separately in this section.

\subsection{The neutrinophilic $\phi_2$}

The relic abundance of dark matter in this scenario is determined by forbidden annihilation processes. While the annihilation channel \(SS \rightarrow N_1 \bar{\nu}_\alpha\) could, in principle, contribute, it is not viable in our setup. For this process to occur, the condition \(M_{N_1} > 2M_{\rm DM}\) must hold. However, this would allow the decay channel \(N_1 \to \nu SS\) to repopulate dark matter and result in overproduction, conflicting with the observed relic density. To prevent this, \(M_{N_1}\) is chosen to be \(< 2M_S\), rendering \(N_1 \to \nu SS\) kinematically forbidden. Additionally, the Yukawa coupling of \(N_1\) is kept sufficiently small, making the \(SS \rightarrow N_1 \bar{\nu}_\alpha\) annihilation channel irrelevant for determining the DM relic density.  

Instead, the forbidden annihilation channel \(SS \rightarrow N_2 \bar{\nu}_\alpha\) is considered to achieve the correct relic density. In this case, the mass of \(N_2\) can be tuned to maintain the dark matter in a kinematically forbidden regime, where \(M_{N_2} > 2M_{\rm DM}\), ensuring the forbidden annihilation process remains efficient. The relative mass difference between the initial and final state particles is parameterized as  
$\Delta = ({M_{N_2} - 2M_{\rm DM}})/{2M_{\rm DM}}$. 
This approach is particularly advantageous in the light dark matter mass region, ranging from 50 MeV to 5 GeV. In this range, achieving the correct relic density for thermal WIMPs is often difficult, and CMB constraints exclude light DM annihilating into SM charged fermions during recombination. Since DM in this scenario annihilates primarily into \(N_2\) and neutrinos, it avoids these constraints while still ensuring the desired relic abundance. Notably, in this scenario, the re-population of DM through the decay $N_2 \to SS \nu$ can be controlled with a suppressed branching fraction to DM as $N_2$ can dominantly decay to $N_1 \nu \nu$.

\begin{figure}[htb]
    \centering
    \includegraphics[scale=0.5]{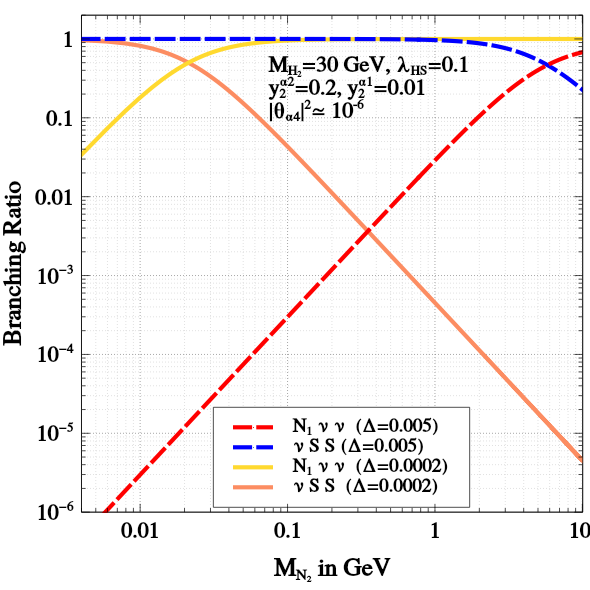}
    \caption{The branching ratios of the two dominant $N_2$ decay channels: $N_2\rightarrow N_1\nu\nu$ and $N_2\rightarrow \nu SS$, for two distinct mass splittings $\Delta$.  }
    \label{brMn2}
\end{figure}

\begin{figure}[htb]
    \centering
    \includegraphics[scale=0.45]{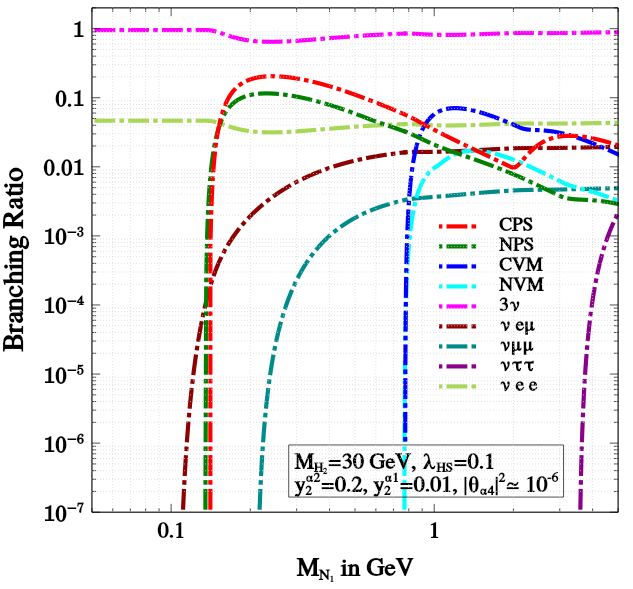}
\hfil\includegraphics[scale=0.45]{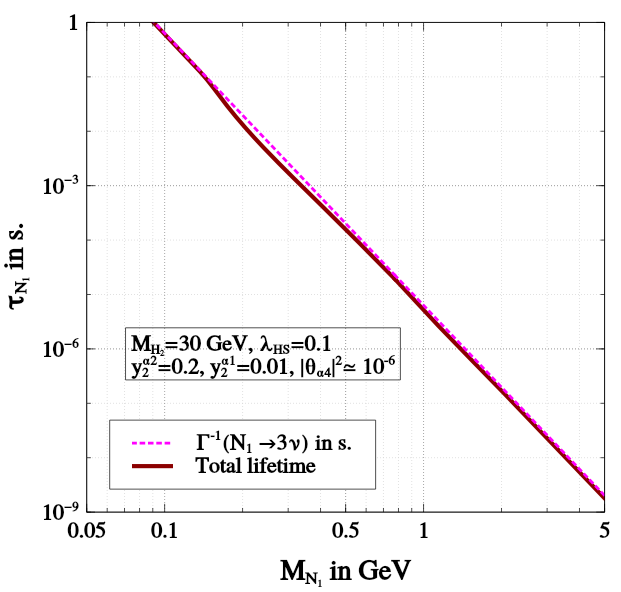}
    \caption{{Branching ratio for various decay modes [left] and lifetime of $N_1$ [right] as a function of $N_1$ mass  ($M_{N_1}$), with other parameters fixed to the values specified in the figure inset. }}
    \label{n1decay}
\end{figure}

Thus the choice of mass splitting $\Delta$ plays a crucial role in this scenario, affecting both the forbidden annihilation process and the potential re-population of DM from $N_2$ decays. To maintain the dominance of the $N_2 \to N_1 \nu \nu$ decay channel and prevent significant DM production through $N_2 \to \nu SS$ decay, a small $\Delta$ is essential. This suppresses the phase space for the $N_2 \to \nu SS$ decay, resulting in a smaller branching fraction.  To illustrate this effect, in Fig.~\ref{brMn2}, we have shown the branching fraction for these two decay channels, as a function of $N_2$ mass for two different values of $\Delta$ {\it i.e.} $\Delta =0.005$ and $\Delta=0.0002$. The results clearly demonstrate that a larger $\Delta$ leads to a higher branching fraction for the $\nu SS$ final state, potentially resulting in DM overproduction. Consequently, an upper bound on $\Delta$ can be imposed, and we will restrict our analysis to a maximum value of $\Delta = 0.01$ for analyzing the allowed parameter space.
Here, it is worth mentioning that $N_1$ undergoes decays into SM final states, thereby depleting its cosmological abundance. The details of the various decay modes of $N_1$ are discussed in the Appendix~\ref{appen1}. The branching ratio to different decay modes as well as the lifetime of $N_1$ are illustrated in the left and right panels of Fig.~\ref{n1decay} respectively.   Fig.~\ref{n1decay} reveals that the dominant decay mode is $N_1 \to 3\nu$, with the branching fraction approaching unity. Moreover, from the right panel, it is clear that for $m_{N_1} \gtrsim 50$ MeV, the lifetime $\tau_{N_1}$ is shorter than 1 second. Crucially, since the decay products consist almost exclusively of neutrinos, this scenario naturally evades constraints from BBN as well as CMB, as the electromagnetic energy injection remains negligible.

The evolution of comoving number densities of DM and $N_2$ can be obtained by solving the Boltzmann equations incorporating all the number changing processes. These equations can be written in terms of comoving number densities as follows.

\begin{eqnarray}
    \frac{dY_S}{dx}&=& - \frac{s}{\mathcal{H} x} \langle \sigma v \rangle_{SS\to N_2 {\nu}} \left[Y^2_S - \left(\frac{Y_{N_2} Y_\nu}{Y^{\rm eq}_{N_2} Y^{\rm eq}_\nu}\right)(Y^{\rm eq}_S)^2\right]\nonumber+\frac{2}{\mathcal{H}x} \mathcal{B}_{N_2\to \nu SS}\langle  \Gamma_{N_2} \rangle \left[Y_{N_2}-\frac{Y^2_S}{(Y^{\rm eq}_S)^2}Y^{\rm eq}_{N_2}\right]\,,\nonumber\\
    \frac{dY_{N_2}}{dx}&=& - \frac{s}{\mathcal{H} x} \langle \sigma v \rangle_{N_2 N_2\to \nu \overline{\nu}} \left[Y^2_{N_2} - (Y^{\rm eq}_{N_2})^2\right]-\frac{s}{\mathcal{H} x} \langle \sigma v \rangle_{N_2 \nu\to SS} \left[Y_{N_2} Y_\nu -\left(\frac{Y^{\rm eq}_{N_2}Y^{\rm eq}_\nu}{(Y^{\rm eq}_S)^2}\right) Y^2_S\right]\nonumber\\&-&\frac{1}{\mathcal{H}x} \langle \Gamma_{N_2} \rangle \left[Y_{N_2}-Y^{\rm eq}_{N_2}\right]\nonumber\\
\end{eqnarray}
where, $\mathcal{H}$ is the Hubble expansion rate of the Universe, $s$ is the entropy density and $x$ is the dimension less parameter $x=M_{\rm DM}/T$. Here, $\Gamma_{N_2}$ is the total decay width of $N_2$ and $\mathcal{B}_{N_2\to \nu SS}$ is its branching fraction into $\nu SS$. {For small values of $\Delta$}, as $N_2 \to  N_1 \nu \nu$ decay is dominant, its branching into $\nu SS$ remains suppressed preventing significant re-population of dark matter through $N_2$ decay.

\noindent The cross-section for the $SS\rightarrow \ N_2 \bar{\nu}_\alpha$ process is
\begin{eqnarray}
    \sigma(SS\rightarrow N_2\bar{\nu}_\alpha)=\frac{v^2y_2^2\lambda_{HS}^2(s-M^2_{N_2})}{32\pi(s-M_{H_2}^2)(s-4M_S^2)} \sqrt{\frac{s-4M_S^2}{s^3}},
\end{eqnarray}
where, $y_2 = \sqrt{\sum_\alpha \lvert y^{\alpha 2}_2 \rvert^2}$.

For the annihilation process in thermal equilibrium, we have  $${(n_{\rm DM}^{\rm eq})^2\langle\sigma v\rangle_{ SS\rightarrow N_2\bar{\nu}_\alpha}=(n_{N_2}^{\rm eq} n^{\rm eq}_\nu)\langle\sigma v\rangle_{ N_2 \bar{\nu}_\alpha\rightarrow SS}},$$ which leads to the relation between the forbidden cross-section and allowed cross-section as
\begin{equation}
    \langle\sigma v\rangle_{ SS\rightarrow N_2\bar{\nu}_\alpha}= ~ \zeta(3) \sqrt{\frac{2}{\pi}}\frac{(M_{N_2} T)^{3/2}}{M^3_S} e^{-2\Delta x} \langle\sigma v\rangle_{ N_2 \bar{\nu}_\alpha\rightarrow SS}
\end{equation}

\begin{figure}[h]
    \centering
    \includegraphics[scale=0.485]{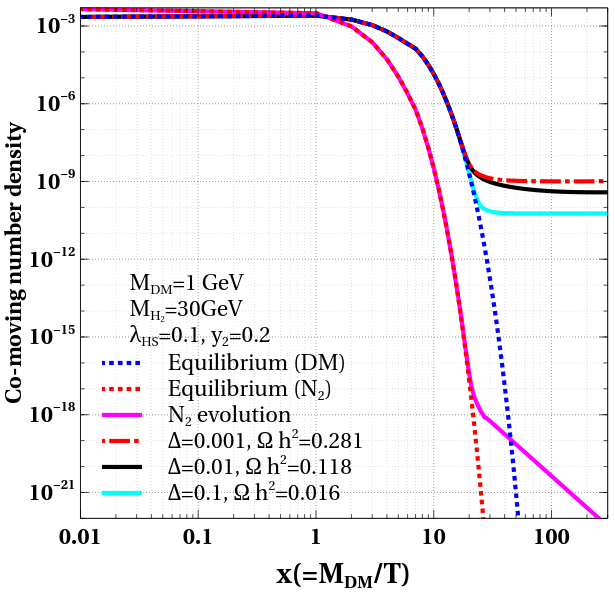}
    \includegraphics[scale=0.485]{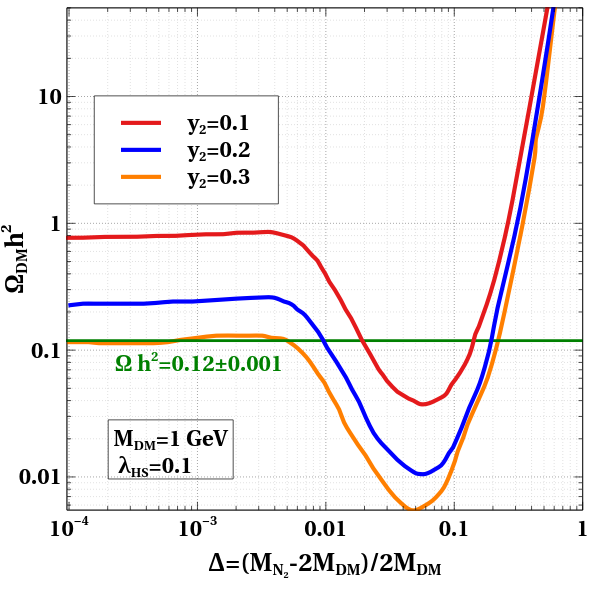}
    \caption{ Evolution of comoving number density of DM and $N_2$. (Right) Variation of relic abundance of DM with $\Delta$ for three benchmark Yukawa couplings.}
    \label{dm1}
\end{figure}
The left panel of Fig. \ref{dm1} illustrates the evolution of comoving abundances for DM and $N_2$ for different values of mass splitting $\Delta$, while keeping other parameters fixed. Clearly, by an appropriate tuning of the couplings and the mass-splitting $\Delta$, it is possible to achieve the correct relic density of DM. The comoving abundance of $N_2$ (magenta line) remains in equilibrium before freezing out with a much smaller abundance compared to DM, indicating that its decay does not significantly affect the abundance of DM post-freeze-out. 

To understand the effect of $\Delta$, the right panel of Fig.~\ref{dm1}, shows the variation of DM relic with $\Delta$ for three different benchmark values of Yukawa coupling $y_2$ keeping other parameters fixed. The variation of relic density demonstrates the role of $\Delta$ in deciding the annihilation rate of DM into the forbidden channel $N_2 \nu$ as well as in determining the branching fraction of $N_2$ to DM which also controls the repopulation of DM. For small values of $\Delta$ (i.e. $\lesssim 0.005$ for this benchmark), we see that relic density remains almost independent of $\Delta$, though there is a marginal rise which is attributed to the modified Boltzmann suppression in the forbidden annihilation cross-section of $SS \to N_2 \nu$, resulting in a smaller annihilation rate and consequently higher relic density. 
However, beyond that value, with increase in $\Delta$, the branching fraction for $N_2 \to \nu SS$ becomes significant. Thus, this decay process, effective during DM freeze-out, enhances the DM number density and delays freeze-out by maintaining the interaction rate above the Hubble expansion rate, even if the cross-section gets some Boltzmann suppression with an increase in $\Delta$. Hence, beyond a certain $\Delta$ threshold, $N_2$ branching to DM become non-negligible, delaying freeze-out and reducing the relic abundance due to an extended interaction period. For even larger $\Delta$, Boltzmann suppression in annihilation rates dominate due to which DM produced from significant decay branching of $N_2 \to \nu S S$ can not annihilate efficiently. This leads to increase in relic abundance again with $\Delta$, as annihilation becomes inefficient. This non-monotonic dependence is clearly visible in the plot.

\begin{figure}[h]
    \centering
    \includegraphics[scale=0.5]{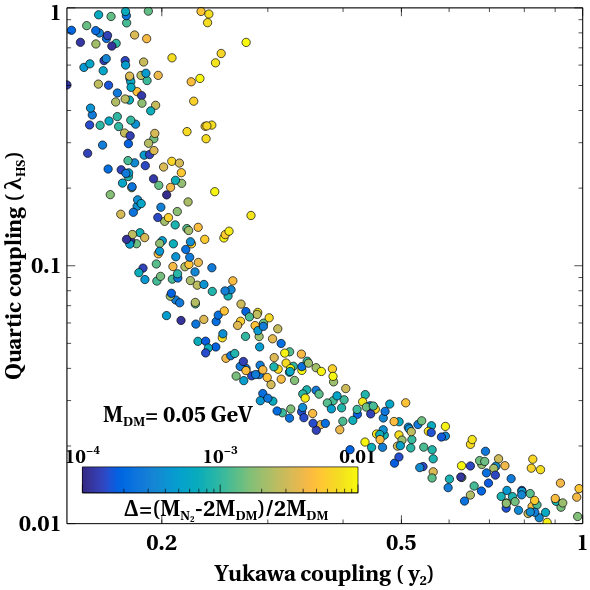}
    \includegraphics[scale=0.5]{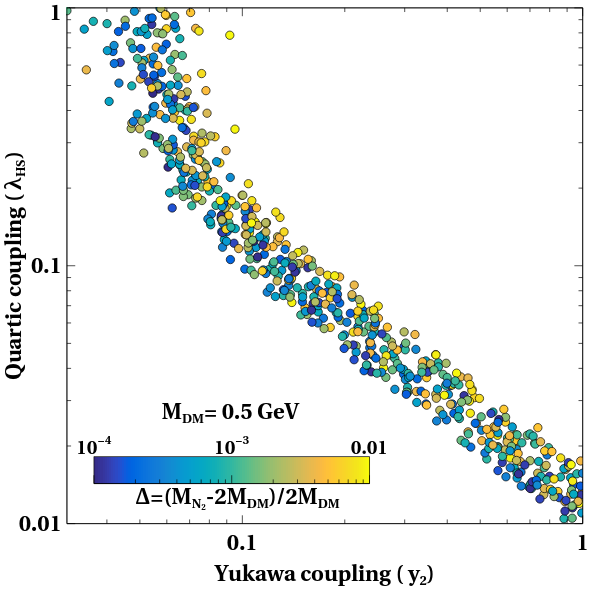}
    \caption{ Relic satisfied points (at 3$\sigma$ C.L.) in Yukawa vs. quartic coupling plane with the color gradient depicting the value of $\Delta$.}
    \label{dm2}
\end{figure}

Fig. \ref{dm2} illustrates the parameter space where the correct DM relic density is achieved, plotted in the $\lambda_{HS}$-$y_2$ plane. We present results for two benchmark DM masses: $M_{\rm DM}=0.05$ GeV (left panel) and $M_{\rm DM}=0.5$ GeV (right panel), with the color gradient indicating different values of $\Delta$. Though correct relic densities are indeed achievable for larger $\Delta$ values, our parameter scan has conservatively been restricted to $\Delta \leq 0.01$.
The plots reveal an inverse relationship between $\lambda_{SH}$ and $y_2$ in achieving the correct relic density. As $\lambda_{SH}$ increases, $y_2$ must decrease to maintain the required DM abundance. 

\subsection{The leptophilic $\phi_2$}

\begin{figure}[h]
    \centering
    \includegraphics[scale=0.47]{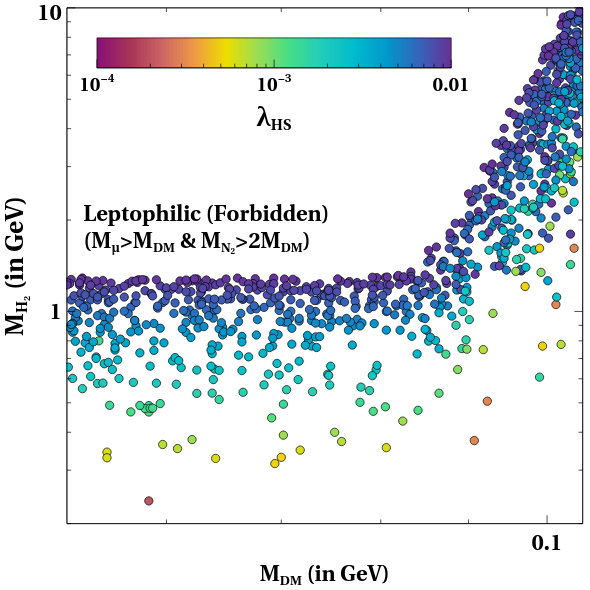}
        \includegraphics[scale=0.47]{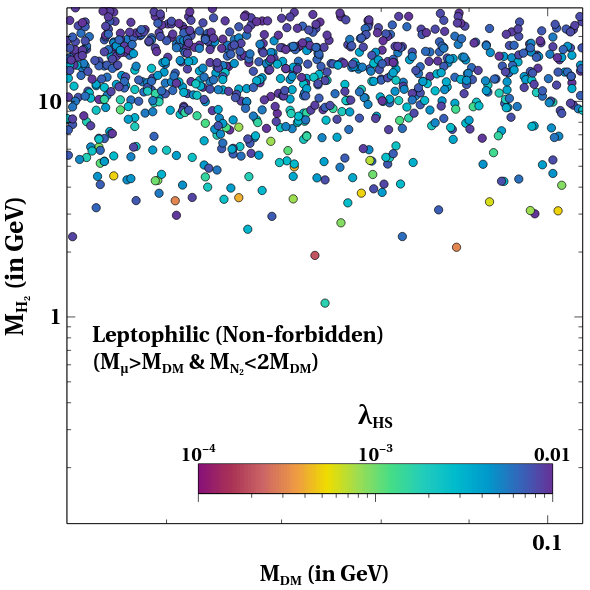}
    \caption{DM relic satisfied points for the leptophilic case with forbidden (left panel) and non-forbidden (right panel) final states. We show the Higgs portal coupling as the continuous colour spectrum here. As we are concerned with the $(g-2)_{\mu}$, therefore, in the leptophilic case, we have scanned up to the mass $\sim 105$ MeV.}
    \label{muonf}
\end{figure}
When we opt for the leptophilic scalar doublet $\phi_2$, it brings about substantial changes in the dark matter parameter space compared to the neutrinophilic scenario. This is due to the fact that the final state masses can not be chosen arbitrarily to keep DM in the forbidden regime. The significance of the neutrino Yukawa coupling ($y_2$) diminishes, while the charged lepton Yukawa coupling ($Y_2$) takes on a more or equally dominant role in determining the relic density. As we consider DM mass range from 50 MeV to a few GeV, we can not keep the $e^+ e^-$ final states forbidden, however, we make it suppressed by choosing electron coupling with $\phi_2$ to be negligible. We keep a sizeable coupling of the muon with $\phi_2$ motivated from $(g-2)_\mu$ and accordingly, keep the $\mu^+ \mu^-$ final states in the forbidden regime. While our primary goal is to study the forbidden DM case with both $N_2 \bar{\nu}_\alpha$ and $\mu^+ \mu^-$ final states in the kinematically forbidden regime, we also check the corresponding results for the non-forbidden case as a comparison. While we still keep the charged lepton final states suppressed or kinematically forbidden, we kinematically allow the $N_2 \bar{\nu}_\alpha$ to go to the non-forbidden regime while still explaining $(g-2)_\mu$ and satisfying CMB bounds on DM annihilation into charged fermion states during recombination.

Fig. \ref{muonf} shows the parameter space in $M_{H_2}-M_{\rm DM}$ plane and scalar portal coupling $\lambda_{HS}$ in colour code, for both forbidden and non-forbidden DM with leptophilic $\phi_2$ as mediator. The Yukawa coupling $Y^{\mu}_2$ is varied in the range consistent with $(g-2)_\mu$ while $Y^e_2, Y^{\tau}_2$ are negligible. The choice of $y_2$ is kept within the interval $ [ 10^{-3}, 10^{-2} ]$. As the left panel of Fig. \ref{muonf} shows, there is an abrupt surge in the required values of $M_{H_2}$, particularly starting from around $M_{\rm DM}\sim 90$ MeV. This is because, as DM mass becomes closer to muon mass, the relative mass splitting $\Delta$ decreases leading to reduced Boltzmann suppression. If we keep the scalar portal and Yukawa couplings fixed, we need to increase the mass of mediator $M_{H_2}$ to get the correct relic abundance while compensating for the decrease in Boltzmann suppression of final states. Similarly, in the right panel for the non-forbidden case, we observe a push for a larger mediator mass to satisfy the relic density constraint while keeping all other parameters fixed. This is once again due to the requirement of reducing the annihilation cross-section to get the correct relic abundance after Boltzmann suppression in final states disappears.
As expected, the non-forbidden process $SS\rightarrow N_2\bar{\nu}_\alpha$ dominates over the forbidden process $SS\rightarrow \mu^+ \mu^-$ in controlling the relic of the non-forbidden scenario.

\begin{figure}[h]
    \centering
    \includegraphics[scale=0.5]{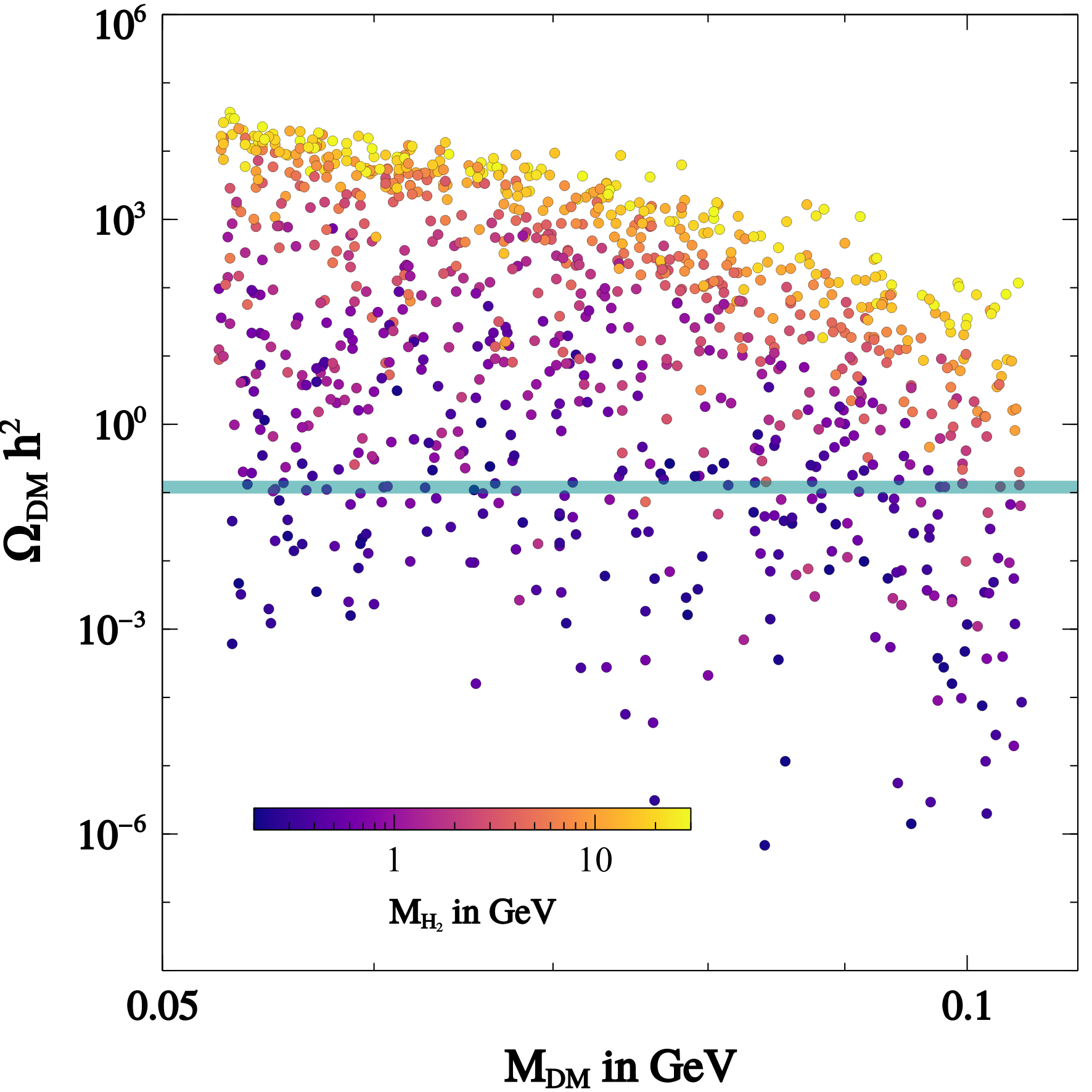}
    \caption{Relic density as a function of DM mass for the points satisfying constraints from $\Delta a_\mu$ and MEG-I. 
    }
    \label{relicLF}
\end{figure} 

Fig.~\ref{relicLF} shows DM relic versus DM mass in the leptophilic $\phi_2$ case with colour code denoting $H_2$ mass. The points shown in this figure satisfy the criteria of $(g-2)_\mu$ and LFV constraints, discussed in section~\ref{sec:g2}. Since the parameters influencing $\Delta a_\mu$ and BR$(\mu\to e\gamma)$ are already restricted due to experimental bounds, the other relevant parameter is varied randomly as $\lambda_{HS} \in[0.001,0.01]$. Evidently, there exists a common parameter space that not only complies with the correct relic density of dark matter but also aligns with the constraints arising from the muon anomalous magnetic moment and LFV.

\subsection{CMB constraints}
DM annihilating into charged fermions or photons during the recombination epoch can cause noticeable distortions in the CMB anisotropy spectrum and hence there exist stringent constraints \cite{Madhavacheril:2013cna,Slatyer:2015jla,Planck:2018vyg}, particularly for light thermal DM of the mass range discussed in this work. Kinematically forbidden DM remains safe from such bounds as DM annihilation to charged fermion states are Boltzmann suppressed at low temperatures.

The neutrinophilic scenario discussed here remains completely safe from such CMB bounds as DM annihilates only into forbidden final states $N_2 \bar{\nu}_\alpha$  or allowed final states $N_1 \bar{\nu}_\alpha$ and even at one-loop level there is no DM annihilation process into photons. However, in the leptophilic case, loop-level annihilation of DM into two photons is still viable and thus imposes a stringent constraint on this annihilation rate during the recombination epoch. Although there is a loop suppression, the final states (photons) are in non-forbidden mode and hence can face tight constraints from CMB. 

\begin{figure}[h]
    \centering
    \includegraphics[scale=0.5]{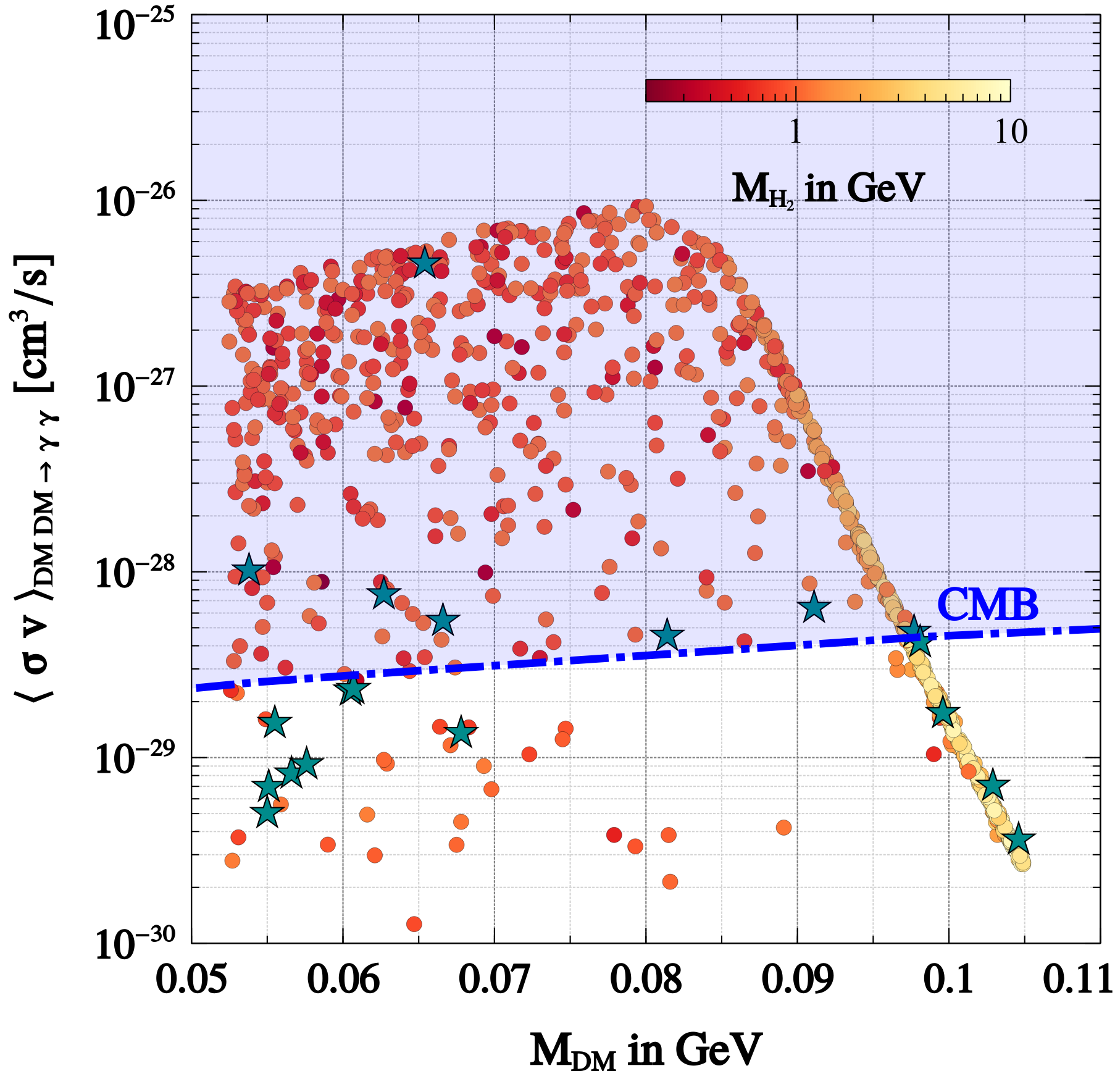}
    \caption{Points satisfying correct relic density, $(g-2)_\mu$ and LFV constraints, against the CMB constraints on $\langle \sigma v \rangle_{{\rm DM DM} \to \gamma \gamma}$. The circular colour-coded points satisfy the correct relic density, while the dark cyan star-shaped points fulfil $(g-2)_\mu$, LFV constraints, and correct relic density simultaneously.}
    \label{fig:cmbcon}
\end{figure}

In Fig.~\ref{fig:cmbcon}, we scrutinize the points satisfying correct relic density as well as $(g-2)_\mu$ and LFV constraints against the constraint on DM annihilation to photons during the CMB decoupling~\cite{Slatyer:2015jla}. The circular colour-coded points, with the colour bar indicating the mass of $M_{H_2}$, correspond to those depicted in Fig. \ref{muonf} meeting the correct relic density constraint for DM. The dark cyan star-shaped points align with the correct relic density satisfying points from Fig.\ref{relicLF}. Consequently, these points collectively fulfil $(g-2)_\mu$, LFV constraints, and correct relic density simultaneously. Clearly, there is a viable parameter space that meets all these criteria while remaining consistent with the  CMB constraints.

Interestingly, even the non-forbidden DM scenario of the model can be saved from the CMB bounds by appropriately choosing the final states. For example, if DM annihilates dominantly into $N_1 \bar{\nu}_\alpha$, the CMB constraints can be made weaker. At first glance, one might assume that this situation is immune to constraints from CMB considerations. However, the accessibility of decay channels for $N_1$, leading to $\nu e^+ e^-$, introduces potential issues, subjecting it to significant constraints imposed by CMB anisotropy bounds. In our specific configuration, the presence of a light scalar $H_2$ results in $N_1$ primarily decaying into $3\nu$, with its decay into $\nu e^+ e^-$ being effectively suppressed. This suppression arises due to the chosen tiny value of $H_2$-electron Yukawa coupling $Y^e_{2}$ and heavy mediator (SM gauge and Higgs bosons) suppression of $N_1 \rightarrow \nu e^+ e^-$ decay channel. Consequently, the entire parameter space manages to evade constraints imposed by CMB considerations in the non-forbidden case.

\subsection{Direct Detection of DM}
\begin{figure}[h]
    \centering
    \includegraphics[scale=0.5]{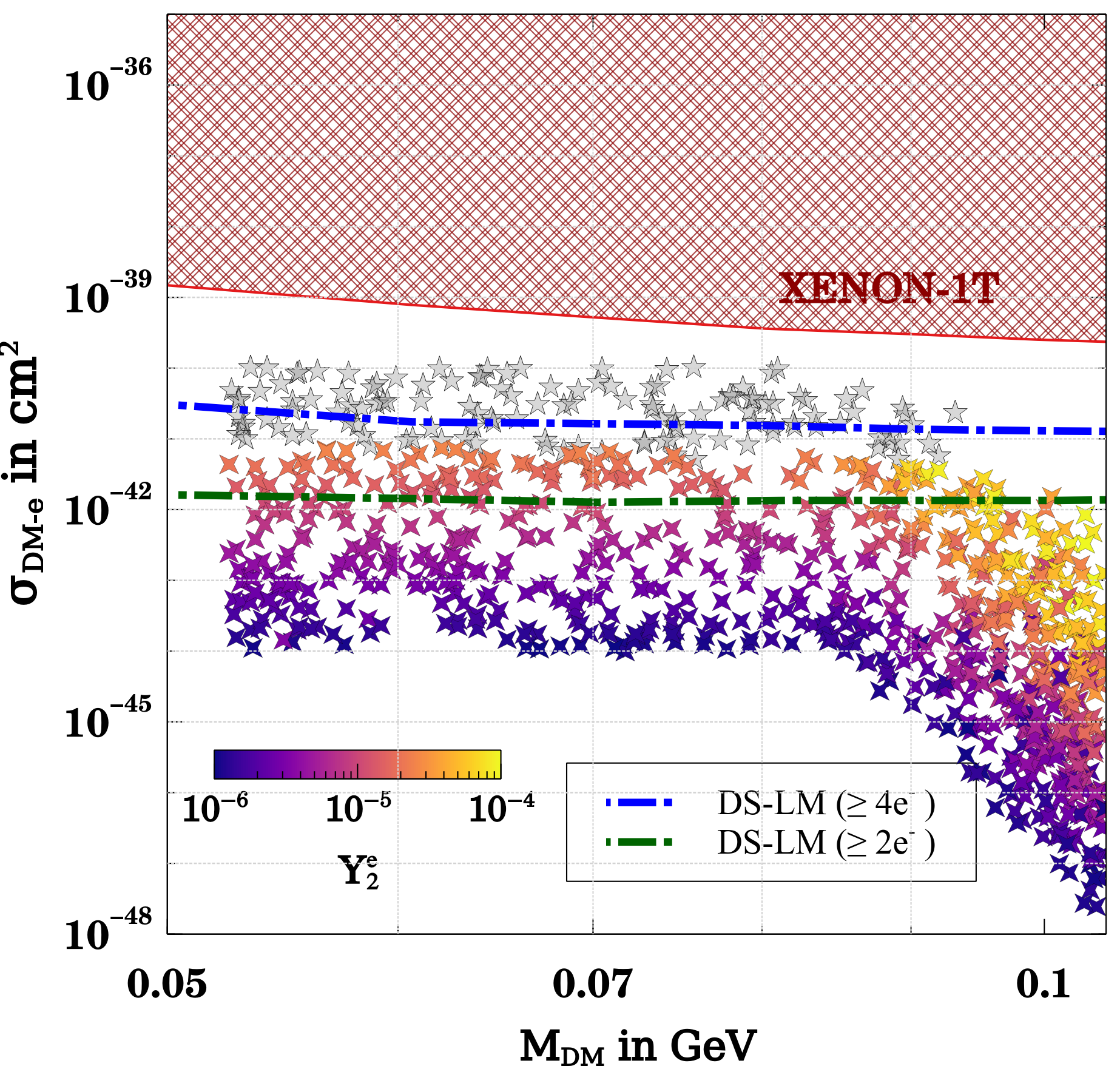}
    \caption{DM-electron scattering cross-section as a function of DM mass for the points satisfying correct relic density. The colour code depicts the value of $Y^e_2$ coupling. The grey points are ruled out by the CMB constraint on DM annihilation into electrons~\cite{Slatyer:2015jla, Planck:2018vyg}. }
    \label{fig:dmescatt}
\end{figure}

In our setup, DM can interact with both nucleons and electrons within terrestrial DM detectors. DM-electron scattering occurs through the $H_2$ scalar in the leptophilic scenario. However, when the coupling between $H_2$ and electrons is substantial, it may pose a challenge, conflicting with constraints derived from CMB observations. This is because the same coupling has the potential to enhance the DM annihilation cross-section into electrons.

In Fig.~\ref{fig:dmescatt}, we showcase the DM-electron scattering cross-section as a function of DM mass for the points satisfying correct relic density as shown in Fig.~\ref{muonf}. The colour code depicts the value of the coupling $Y^e_2$ which we vary in a range $[10^{-6},10^{-4}]$. The grey-coloured points are ruled out by the constraint on DM annihilation to electrons from CMB~\cite{Slatyer:2015jla, Planck:2018vyg}. We also showcase the most stringent constraints and the on DM-electron scattering cross-section from XENON-1T~\cite{XENON:2019gfn} and the projected sensitivity of DS-LM~\cite{GlobalArgonDarkMatter:2022ppc}. The shaded regions depict the constraint from XENON-1T and the dot-dashed lines depict the projected sensitivity of DS-LM.

\begin{figure}[h]
    \centering
    \includegraphics[scale=0.5]{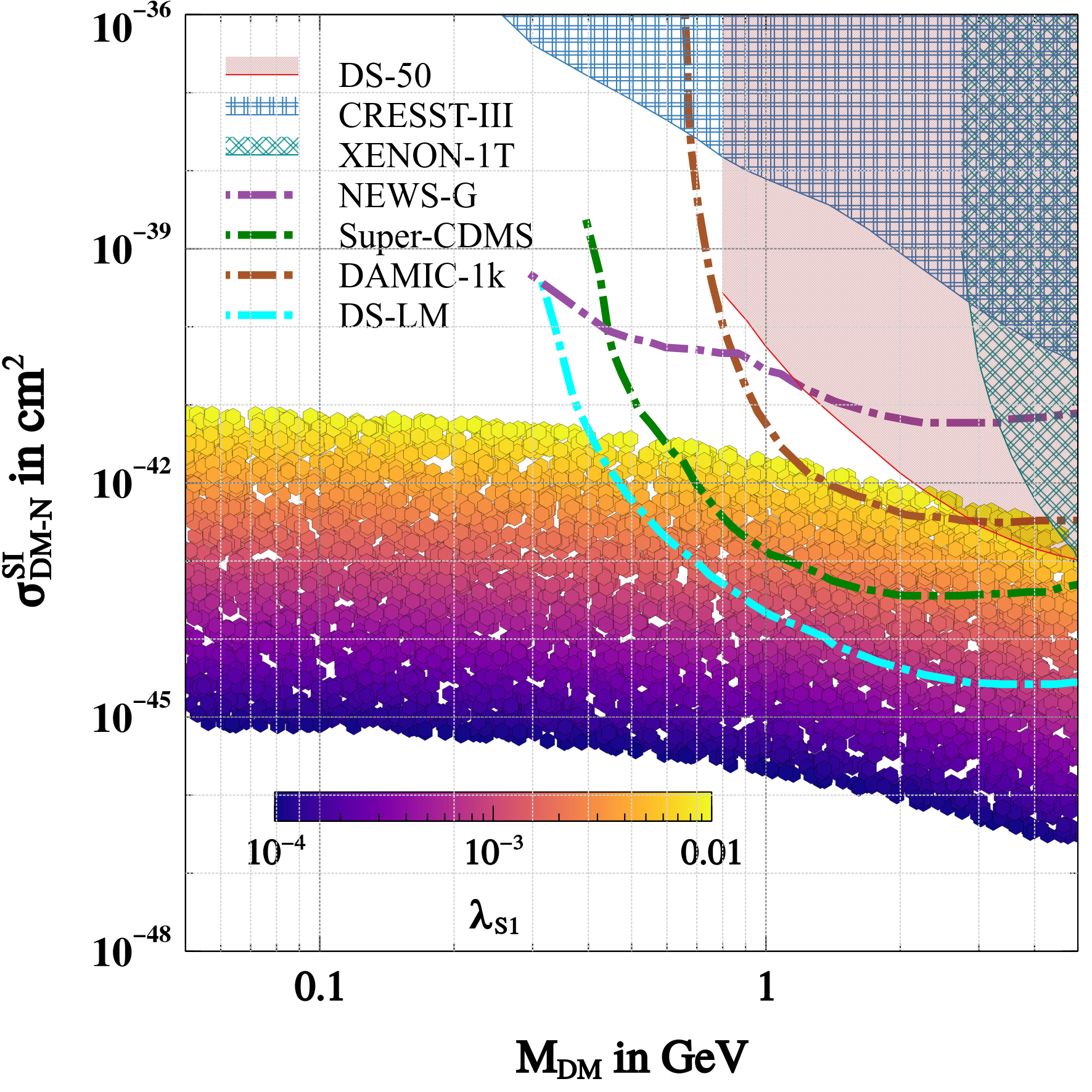}
    \caption{Spin-independent DM-nucleon scattering as a function of DM with the $\lambda_{S1}$ coupling shown in the colour code. }
    \label{fig:dmnscatt}
\end{figure}

The DM-nucleon scattering can take place via the SM Higgs mediation. In Fig.~\ref{fig:dmnscatt}, we showcase the spin-independent DM-nucleon scattering cross-section as a function of DM mass by varying the coupling $\lambda_{S1}$ in a range $[10^{-4},10^{-2}]$ which is depicted in the colour code. We also project the constraints from XENON-1T~\cite{XENON:2019gfn}, CRESST-III~\cite{CRESST:2019jnq} and DS-50~\cite{DarkSide-50:2022qzh} by the shaded regions as well as the projected sensitivities of NEWS-G, Super-CDMS~\cite{SuperCDMS:2016wui}, DAMIC-1k~\cite{DAMIC:2019dcn} and DS-LM~\cite{GlobalArgonDarkMatter:2022ppc} by the dot-dashed lines. We also check that the chosen values of $Y^e_2, \lambda_{S1}$ in the direct-detection analysis do not affect the relic abundance of DM.

\section{Detection Prospects of HNL}
\label{sec6}
HNL with masses in the MeV to GeV scale has compelling detection prospects in present and future target experiments as they have been searched for via the signature of kinks in the Kurie plots in nuclear beta decays, via anomalous peaks in the energy spectra of charged leptons in two-body leptonic decays of pseudoscalar mesons and in the apparent deviation of ratio of branching fraction of mesons to leptons from their SM values. For example, HNL can lead to a deviation of the ratio BR$(K^+ \to e^+ +\nu_e)/ {\rm BR}(K^+\to\mu^+ +\nu_\mu)$ as well as for the decay of $\pi^+$. Similarly, HNL can be probed in the apparent deviation of the spectral parameters in $\mu$ and leptonic $\tau$ decay from their SM values. Here it is worth mentioning that the HNL masses and mixing with distinct neutrino flavours can be considered independent free parameters from a model-independent standpoint. However, generating light neutrino masses imposes theoretical restrictions, creating a link between the HNL and active neutrino sectors that might be used as guidance for future experimental studies.

HNL production and decay in minimal scenarios are governed by SM interactions and the mixing of HNL with the active neutrino, resulting in relatively long lifetimes if the masses are in the MeV-GeV range. This is the foundation of searches, such as those conducted at colliders and beam dump experiments. HNL have new sources of production and decay channels in models with more interactions. If the extra interactions associated with dark matter become stronger, it may increase scattering cross sections and quick decays, fundamentally altering the HNL phenomenology. If it decays promptly into neutrinos or other unseen particles, for example, it can weaken collider and beam dump limitations since the HNL would not have reached the detector, and other signals would have been reduced by the branching ratio into invisible channels. In our scenario, though HNL has additional interaction, if $H_2$ is heavier then the standard searches for HNL remain valid. However, if $H_2$ is lighter than HNL,  then $N_1$ will dominantly decay to $\nu$ and $H_2$ and hence all these constraints become irrelevant in that scenario.

\begin{figure}[h]
    \centering
    \includegraphics[scale=0.5]{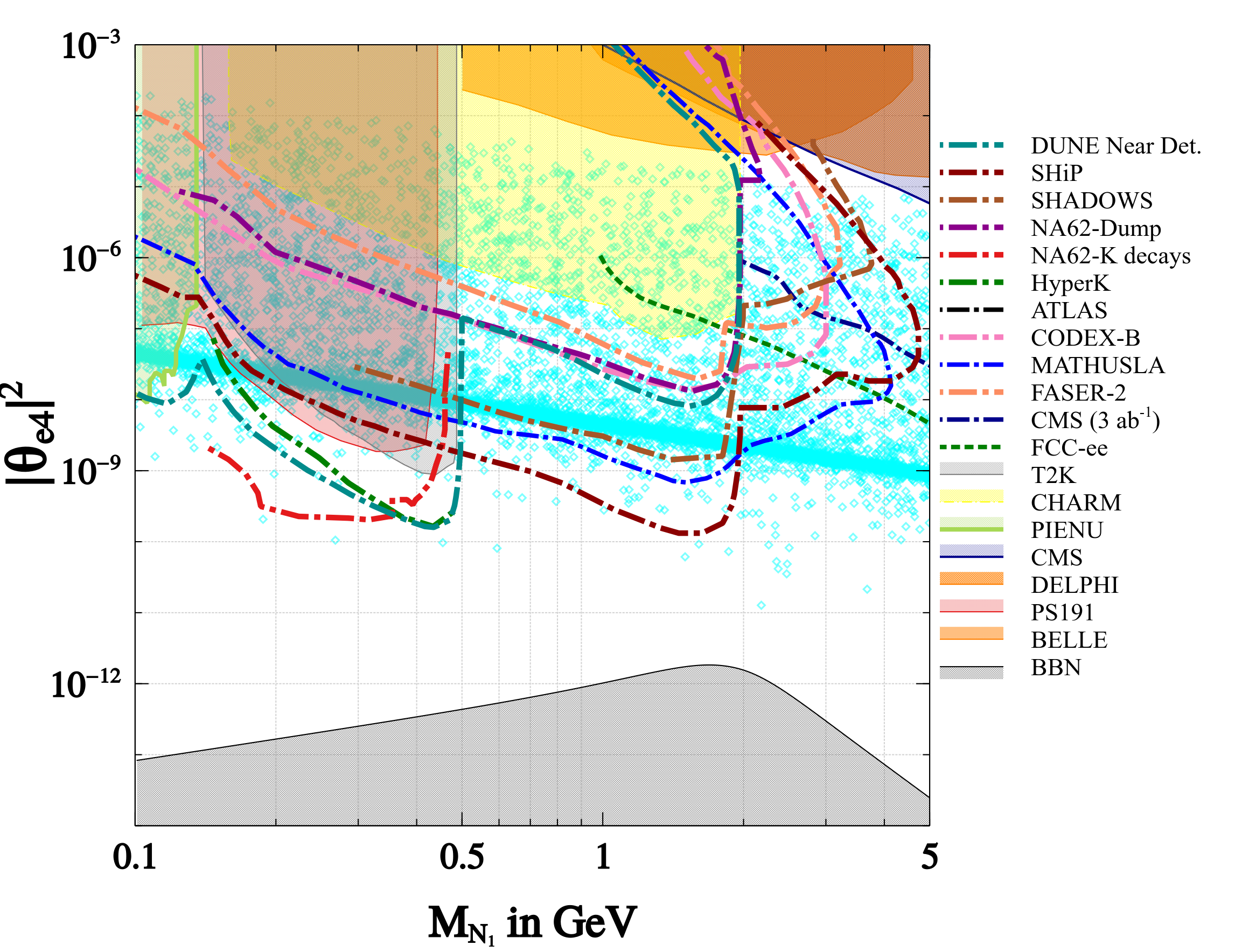}
    \caption{$|\theta_{e4}|^2$ as a function of $M_{N_1}$ in our setup is shown by the cyan-coloured points. The shaded regions depict the existing constraints and the dot-dashed, dashed lines depict the projected sensitivities of various experiments. The gray region at the bottom depict the region excluded by BBN constraints on HNL lifetime.}
    \label{hnl_search}
\end{figure}

In  peak search experiments, which specifically search for anomalous monochromatic peaks in the charged lepton spectra from meson decays, the best constraints on HNL are from $\tau$ lepton and meson decays. Some of the considered meson decays for HNL searches are $\pi^{\pm} \to \ell^{\pm}N_1$, $K^{\pm} \to \ell^{\pm}N_1$, $K^{\pm} \to \pi^0 \ell^{\pm}N_1$,$D^{\pm} \to \ell^{\pm}N_1$,$D^{\pm}_s \to \ell^{\pm}N_1$, $D^{+} \to \ell^{+} \overline{K^0} N_1$ and its conjugate decay and $\tau$ lepton decays $\tau \to \pi N_1$, $\tau \to \rho N_1$ and $\tau \to \ell ~\bar{\nu}_l N_1$~\cite{Abdullahi:2022jlv,Antel:2023hkf}. Even higher masses can be probed via $B$ meson decays $B \to X N_1$, where $X$ are mesons. Similarly, a light HNL can be regarded as a long-lived particle (LLP) in beam dump experiments. Mesons produced during particle accelerator beam collision events can decay to HNL and SM particles. The long-lived HNL can then travel away from the beam collision zone without being affected and decay elsewhere in the detector. At the LHC, HNL can be produced in the GeV mass range through heavy meson decays, $\tau$ leptons, $W$ and $Z$ bosons, Higgs bosons, and even top quarks. Above $B$ meson mass, HNL can be investigated in high luminosity LHC (HL-LHC) by searching for displaced vertices. Also, the HNL being lighter than $Z$ boson mass can also be produced by $Z \to \nu N_1$. This can be probed via a proposed FCC-ee experiment\cite{FCC:2018evy}.

In Fig.~\ref{hnl_search}, the values of active neutrino-HNL mixing angle $\theta_{e4}$ calculated using Casas-Ibarra parameterization in the type-I seesaw scenario (given in Eq. \eqref{ciparam}) are shown via the cyan points where we have imposed the upper limit on absolute neutrino mass from the KATRIN experiment~\cite{KATRIN:2021uub}. 
Existing limits from PS191~\cite{Bernardi:1987ek}, CHARM~\cite{CHARM:1985nku}, PIENU\cite{PIENU:2017wbj}, 
BEBC~\cite{Barouki:2022bkt}, NA62\cite{NA62:2020mcv}, T2K\cite{T2K:2019jwa}, Belle\cite{Belle:2013ytx}, DELPHI\cite{DELPHI:1996qcc}, ATLAS\cite{ATLAS:2019kpx}, and CMS~\cite{CMS:2022fut} are shown by differently shaded regions and the projected sensitivities from the NA62- dump\cite{Beacham:2019nyx}, NA62 $K^+$ decays\cite{NA62:2020mcv}, SHADOWS\cite{Baldini:2021hfw}, SHiP\cite{SHiP:2018xqw}, DUNE near detector\cite{Breitbach:2021gvv}),Hyper-K\cite{T2K:2019jwa}, FASER\cite{FASER:2018eoc}, Codex-b\cite{Aielli:2019ivi} and MATHUSLA\cite{Curtin:2018mvb} are shown by the dot dashed or dotted lines, as indicated by the labels. The BBN lower limit on the mixing parameter (or upper limit on HNL lifetime) ensures that HNL decay does not affect the successful BBN predictions. This constraint has been depicted by the gray shaded region at the bottom of Fig. \ref{hnl_search}. We have considered a viable parameter space for the HNL mass from 100 MeV to 10 GeV from the requirement of achieving the correct relic density. Discussions on even lower HNL mass can be found in \cite{Ruchayskiy:2011aa, Ruchayskiy:2012si}. To be consistent with the constraints from BBN, we restrict the HNL lifetime to be less than $0.1$ s. For this, we take into account all the decay channels of RHN and calculate its total decay width.  With an increasing mass of the RHN, new decay channels open up. Depending on the final state particles, these decay channels can be divided into semileptonic (hadronic) and purely leptonic processes. The details of the lifetime calculation are given in appendix \ref{appen1}. We consider $M_{H_2}=30$ GeV, $\lambda_{HS}=0.1$ and $y^{\alpha 1}_2=10^{-3}$ considering $\phi_2$ to be neutrinophilic in these calculations. The type-I seesaw is assumed to contribute approximately $50\%$ to light neutrino mass throughout our analysis.

\begin{figure}[h]
    \centering
    \includegraphics[scale=0.5]{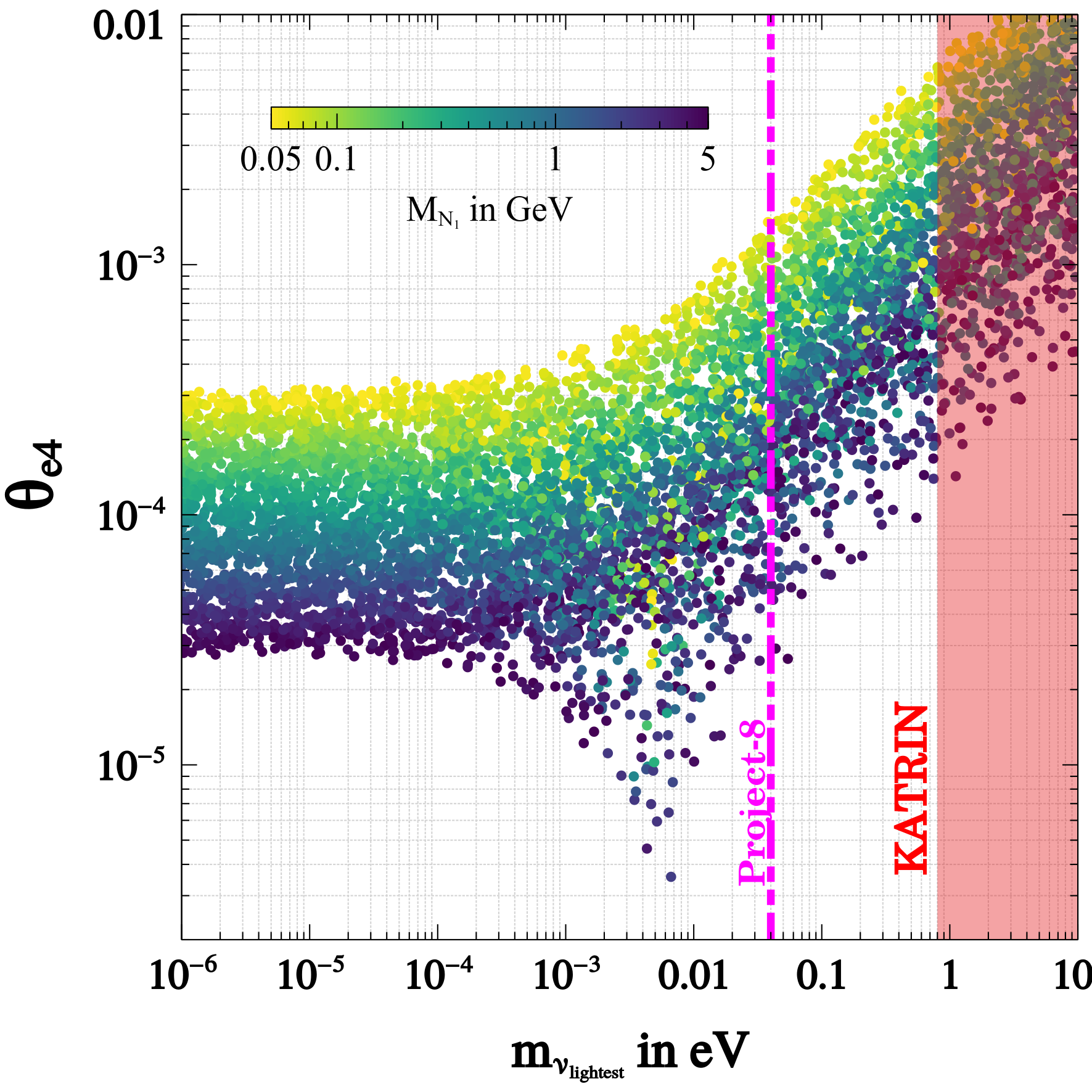}
    \caption{HNL-SM neutrino mixing angle $\theta_{e4}$ as a function of lightest neutrino mass.}
    \label{fig:thetam1}
\end{figure}

Fig. \ref{fig:thetam1} shows the active neutrino-HNL mixing angle as a function of the lightest active neutrino mass, with the colour code depicting the HNL mass. This is generated by randomly varying the complex orthogonal matrix angles $z_{ij}$, used in Casas-Ibarra parameterisation (Eq. \eqref{ciparam}) in $(0-2\pi)$ range for both real and imaginary parts. 
In addition to cosmological constraints on HNL mentioned above, the light neutrino mass is also constrained by cosmology as well as terrestrial experiments. In addition to existing PLANCK constraints on sum of absolute neutrino mass $\sum_i m_i < 0.12$ eV \cite{Planck:2018vyg}, near future CMB experiments like CMB-S4~\cite{CMB-S4:2022ght}, galaxy surveys like DESI~\cite{DESI:2016fyo} or Euclid~\cite{Audren:2012vy} can do further scrutiny. While we show the terrestrial experiment KATRIN bound \cite{KATRIN:2021uub} by the shaded region in Fig. \ref{fig:thetam1}, future sensitivity of other laboratory experiments Project 8~\cite{Project8:2017nal} is shown as the pink dashed line.

\section{Conclusion}
\label{sec7}
We have studied the possibility of light thermal dark matter from sub-GeV to GeV scale by considering a type-I seesaw scenario extended with a second Higgs doublet $\phi_2$. Light DM, assumed to be a real scalar singlet, can have efficient annihilation rates mediated by the neutral component of $\phi_2$, denoted by $H_2$. While $\phi_2$ does not couple to quarks, it can couple to RHN and SM leptons. We consider the alignment limit where the neutral component of $\phi_2$ does not acquire any VEV. However, $\phi_2$ coupling with RHN and SM leptons can still give rise to a sizeable contribution to light neutrino masses via the one-loop effect. Depending upon its neutrinophilic or leptophilic nature of $H_2$, we study the DM phenomenology by considering equal contribution from tree-level and one-loop seesaw to light neutrino mass. 

We study the possibility of kinematically forbidden final states which help in avoiding stringent CMB bounds on light DM annihilation to charged fermion states or photons during recombination. The neutrinophilic scenario is considered in the forbidden mode as the heavy RHN final states can subsequently decay into charged leptons and photons inviting stringent constraints. While a complete study of indirect detection and CMB constraints on DM annihilation into RHN-SM neutrino states in non-forbidden or allowed regime is beyond the scope of the present work, we adopt a conservative approach to keep such final states in forbidden mode for the neutrinophilic case. The leptophilic scenario, disfavoured in the non-forbidden mode, faces tight constraints in the forbidden mode as well due to the one-loop annihilation rate into photons. Motivated by explaining the anomalous magnetic moment of the muon, we consider muon final states in kinematically forbidden regime in the leptophilic scenario while other charged fermion final states do not arise due to suppressed couplings with $\phi_2$. In the leptophilic scenario, we also show the distinct features in terms of parameter space if we keep the muon final states in the forbidden 
regime while kinematically allowing RHN and SM neutrino final states. The model not only explains muon anomalous magnetic moment and CDF-II W-mass anomaly but can also saturate the charged lepton flavour violation limits. Opening up SM Higgs portal coupling of DM or allowing 
 $\phi_2$ coupling to electrons, without contributing significantly to relic while being safe from CMB bounds, can give rise to tree-level direct-detection prospects for DM.

This keeps the parameter space within the reach of several direct detection experiments sensitive to light DM. We also discuss the tantalising detection prospects of RHN whose mass remains in the same range as DM mass in order to appear in final states, either in forbidden or non-forbidden mode.

\section*{Acknowledgements}
 The work of DB is supported by the Science and Engineering Research Board (SERB), Government of India grant MTR/2022/000575. PD would like to acknowledge IITG for the financial support under the project grant number: IITG/R$\&$D/IPDF/2021-22/20210911916. SM acknowledges the financial support from the National Research Foundation of Korea grant 2022R1A2C1005050. The work of NS is supported by the Department of Atomic Energy-Board of Research in Nuclear Sciences, Government of India (Ref. Number:  58/ 14/ 15/ 2021- BRNS/ 37220).

\appendix
\section{Decay modes of HNL}
    \label{appen1}

\begin{figure}[h]
    \centering
    \includegraphics[scale=0.5]{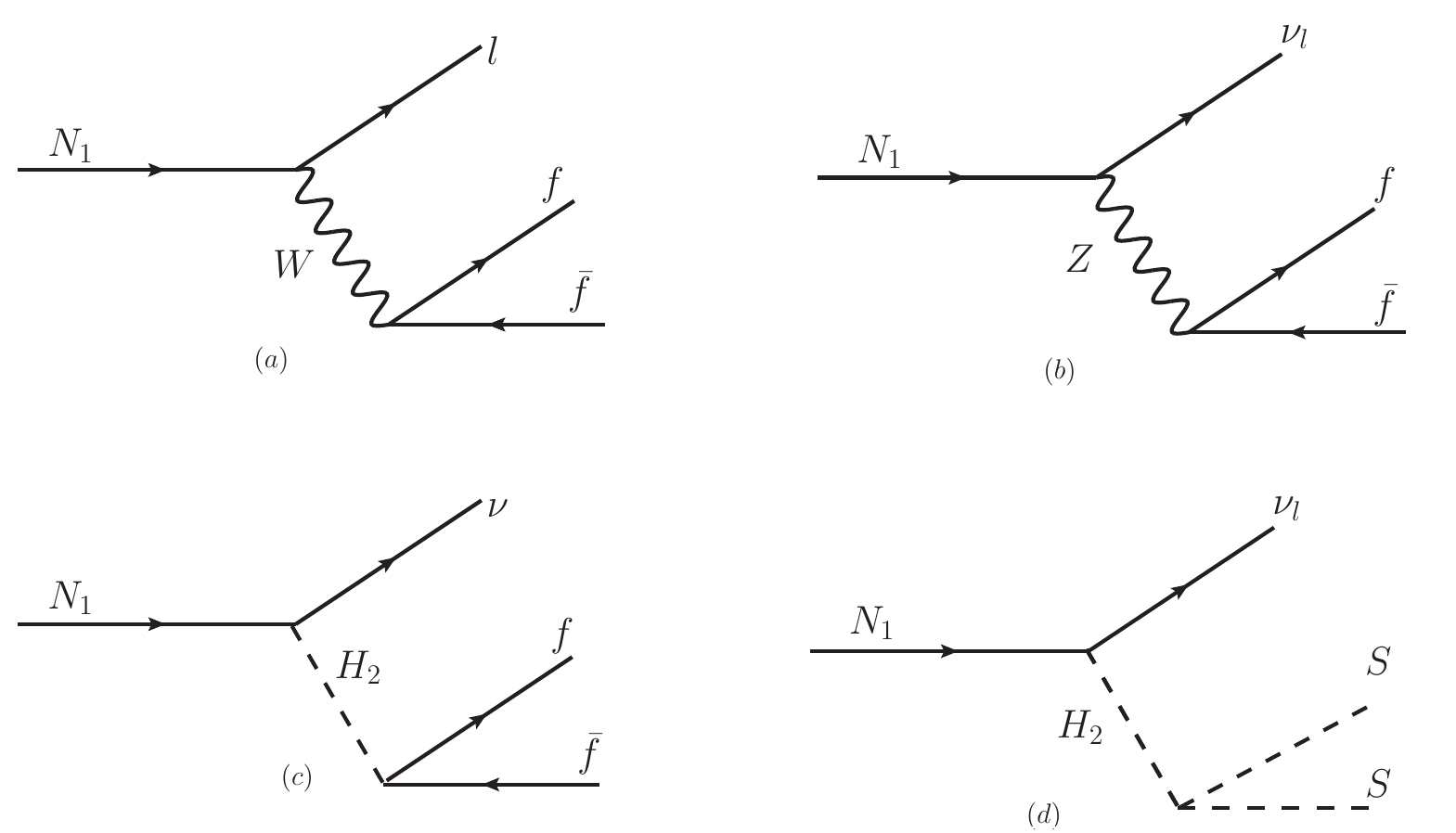}
    \caption{Two-body and three-body decay modes of HNL.}
    \label{nrdecay}
\end{figure}
\subsection{Decay into hadrons}
\label{sec:decay-into-hadrons}

In this section, we briefly discuss the decay modes of the
heavy Majorana neutrino $N_1$, with mass $M_{N_1}$, much smaller than
the mass of the W boson, $M_W$. 

The charged current and neutral current vertices of $N_1$ with the mixing elements are given in
Fig.~\ref{nrdecay} .

In this section, we consider hadronic final states for $M_{N}$ both below and above $\Lambda_{\rm QCD}$ scale. 
%and discuss the range of validity of our results.
The quark pair predominantly binds into a single meson at $M_{N} \lesssim \Lambda_{\rm QCD}$.
There are charged current and neutral current mediated processes with a meson in the final state: $N\to \ell_{\alpha} h_{P/V}^+$ and $N\to \nu_{\alpha} h_{P/V}^0$, where $h_{P}^+$ ($h_P^0$) are charged (neutral) pseudoscalar mesons and $h_V^+$ ($h_V^0$) are charged (neutral) vector mesons.
In formulas below $x_i \equiv m_i / M_{N}$ (with $i$ being the respective particle) , $f_h$ and $g_h$ are the corresponding meson decay constants (see \cite{Bondarenko:2018ptm} for all the numerical values), $\theta_W$ is the Weinberg angle. We have considered only those final state particles whose masses are below 10 GeV, such that $N$ decay to them is kinematically allowed.
The details of the calculations can be found in \cite{Bondarenko:2018ptm, Atre:2009rg}.
\begin{itemize}
    \item The decay width to the charged pseudo-scalar (CPS) mesons ($P^+=\pi^\pm,K^\pm, D^\pm, D_s, B^\pm, B_c$) is given by
\begin{align}
\Gamma^{CPS}	\equiv\Gamma(N\to \ell_{\alpha}^- P^+) &= 
	\frac{G_F^2 f_p^2 |V_{UD}|^2 |\theta_{\alpha 4}|^2 M_{N_1}^3}{16\pi} 
	\left[ \left( 1 - x_\ell^2 \right)^2 - 
	x_P^2(1 + x_\ell^2) \right] 
	\sqrt{\lambda(1, x_P^2, x_\ell^2)},\label{eq:17}
\end{align}
Here $\lambda$ being the Kallen function \cite{Kallen:1964lxa}, defined as: \bea\lambda(x,y,z)=x^2+y^2+z^2-2xy-2yz-2zx.\eea
\item The decay width to the neutral pseudo-scalar (NPS) meson ($P^0=\pi^0,\eta,\eta',\eta_c$) is given by
\begin{align}
\Gamma^{NPS} \equiv \Gamma(N\to \nu_{\alpha} P^0) &=
  \frac{G_F^2 f_P^2 M_{N_1}^3}{32\pi} |\theta_{\alpha 4}|^2 
  \left( 1 - x_P^2 \right)^2
    \label{eq:29}
\end{align}
\item The HNL decay width into charged vector mesons (CVM) ($V^+=\rho^{\pm}$, $a_1^{\pm}$, $D^{\pm*}$, $D^{\pm*}_s$) is given by
\begin{align}
\Gamma^{CVM} \equiv  \Gamma(N\to \ell_{\alpha}^- V^+)= &\nonumber\frac{G_F^2 g_V^2 |V_{UD}|^2 |\theta_{\alpha 4}|^2 M_{N}^3}{16\pi m_V^2}  
	\left(
	\left(1 - x_\ell^2\right)^2 + x_V^2 \left(1 + x_\ell^2\right) - 2 x_V^4
	\right)\\&\times \sqrt{\lambda(1, x_V^2, x_\ell^2)}
	\label{eq:40}
\end{align}
\item 
For the decay into neutral vector meson (NVM) ($V^0=\rho^0$, $a_1^0$, $\omega$,
$\phi$, $J/\psi$) we found that the result depends on the quark content of the meson. To consider it, a dimensionless parameter $\kappa_h$ is introduced, factor to the meson decay constant~\cite{Bondarenko:2018ptm}.
The decay width is given by
\begin{equation}
\Gamma^{NVM}\equiv \Gamma(N\to \nu_{\alpha} V^0) = \frac{G_F^2 \kappa_h^2 g_h^2 |\theta_{\alpha 4}|^2  M_{N}^3}{32\pi m_V^2}
 \left(1 + 2 x_V^2\right) \left(1 - x_V^2\right)^2.
 \label{eq:41}
\end{equation}
\item $N \rightarrow \ell^-_1 \ell^+_2 \nu_{\ell_2}$ where $\ell_1, \ell_2
= e, \mu, \tau$ with $\ell_1 \ne \ell_2$. This decay mode has
charged current interactions only and the decay width is given by
\bea \label{app3bodycc} 
\Gamma^{\ell_1 \ell_2 \nu_{\ell_2}} &\equiv& \Gamma(N \rightarrow \ell^-_1 \ell^+_2 \nu_{\ell_2}) = \frac{G^2_F}{192 \pi^3} M_{N}^5\ {|\theta_{\alpha 4}|}^2\ I_1 (x_{\ell_1},x_{\nu_{\ell_2}},x_{\ell_2}),\\
\nonumber\text{with }I_1(x,y,z) &=& 12 \int\limits_{(x+y)^2}^{(1-z)^2}
\frac{ds}{s}(s-x^2-y^2)(1+z^2-s)\lambda^{\frac{1}{2}}(s,x^2,y^2)
\lambda^{\frac{1}{2}}(1,s,z^2), \eea
where $I_1(0,0,0)=1$. We have set the mass of the light neutrino to zero with a very good approximation in the expression for the width above and henceforth.

\item  $N \rightarrow \nu_{\ell_1} \ell^-_2 \ell^+_2 $ where $\ell_1, \ell_2 = e, \mu, \tau$. Both charged current and
neutral current interactions are relevant for this mode and the decay width is given by
\bea
\label{app3bodyNC1}
\nonumber
\Gamma^{\nu_{\ell_1} \ell_2 \ell_2}&\equiv& \Gamma(N \rightarrow \nu_{\ell_1} \ell^-_2 \ell^+_2 )= \frac {G^2_F}{96 \pi^3}{|\theta_{\alpha 4}|}^2\ M_{N}^5 \times \biggl [ \Bigl ( g^\ell_L g^\ell_R +  \delta_{\ell_1 \ell_2} g^\ell_R \Bigr ) I_2(x_{\nu_{\ell_1}}, x_{\ell_2},  x_{\ell_2}) \\
&+& \Bigl ( {(g^{\ell}_L)}^2 + {(g^{\ell}_R)}^2  + \delta_{\ell_1 \ell_2} (1 + 2g^\ell_L ) \Bigr )I_1(x_{\nu_{\ell_1}}, x_{\ell_2},  x_{\ell_2}) \biggr ] ,\\
\nonumber\text{with, }I_2(x,y,z) &=& 24yz \int\limits_{(y+z)^2}^{(1-x)^2}
\frac{ds}{s}(1+x^2-s)\lambda^{\frac{1}{2}}(s,y^2,z^2)\lambda^{\frac{1}{2}}(1,s,x^2),
\eea
where $I_2(0,0,0) = 1$,
$g^\ell_L = -\frac{1}{2} + x_w$, $g^\ell_R = x_w$ and $x_w = \sin^2\theta_w = 0.231$, where $\theta_W$
is the Weinberg angle.

\item  $N \rightarrow \nu_{\ell_1} \nu \overline{\nu} $
where $\nu_{\ell_1} = \nu_e, \nu_\mu, \nu_\tau$. This decay mode  has a neutral current
interactions only. Using the
massless approximation for the neutrinos as
described above
the decay width has a simple form given by

\bea \label{app3bodyNC2} \Gamma^{3\nu} \equiv
\sum_{\ell_2=e}^\tau \Gamma(N \rightarrow \nu_{\ell_1} \nu_{\ell_2}
\overline{\nu_{\ell_2}} )= \frac {G^2_F}{96\pi^3}|\theta_{\alpha 4}|^2\ M_{N}^5. \eea
\item  The decay width of this tree level decay process, when $M_{N}>M_{H_2}$ is given by:
\bea
\Gamma^{H_2\nu}=\frac{1}{8\pi}y_2^2M_{N}.
\eea

The second Higgs-mediated $N\rightarrow3\nu$ process decay width is given by,
\bea
\Gamma^{3\nu}_{H_2}=\frac{y_2^4 M_{N}^5}{64\pi^3M_{H_2}^4}|\theta_{\alpha 4}|^2,
\eea
and for the $N\rightarrow e\nu_2\nu_2$ decay, the decay width is,
\bea
\Gamma_{H_2}^{e\bar{e}\nu}=\frac{y_2^2(Y_2^{e1})^2M_{N}^5}{64\pi^3M_{H_2}^4}.
\eea
\item $N\rightarrow \nu_l SS$ mediated via $H_2$, when $M_{N}<M_{H_2}$.
\bea
\Gamma^{\nu_lSS}_{H_2}=\sum_{l=e}^{\tau}\Gamma(N\rightarrow \nu_l SS)=\frac{y_2^2\lambda_{HS}^2v^2}{64\pi^3 M_{H_2}^4}|\theta_{l4}|^2M_{N}^3[I_2(0,x_S,x_S)+2I_1(0,x_S,x_S)]\nonumber\\
\eea
\end{itemize}

All the decay modes listed above contribute to the total decay
width of the heavy Majorana neutrino which is given by:
\bea \label{apptotwid} \nonumber
\Gamma_{N}^{\rm Total}& =& \sum_{\ell, P}{\Gamma^{CPS}} + \sum_{\ell, V} {\Gamma^{NPS }} + \sum_{\ell,P} {2 \Gamma^{CVM}} +  \sum_{\ell,V} {2 \Gamma^{NVM}}
+  \sum_{\ell_1,\ell_2(\ell_1 \ne \ell_2)}{2\Gamma^{\ell_1
\ell_2 \nu_{\ell_2}}} \\&&+ \sum_{\ell_1, \ell_2} {\Gamma^{\nu_{\ell_1} \ell_2 \ell_2}} +  \sum_{\nu_{\ell_1}}
{\Gamma^{3\nu}}+ \sum\Gamma^{3\nu}_{H_2}+\sum\Gamma^{e\bar{e}\nu}_{H_2}+\sum\Gamma_{H_2}^{\nu_lSS},
 \eea
where $\ell, \ell_1, \ell_2 = e, \mu, \tau$. For a
Majorana neutrino, the $\Delta L = 0$ process $N \rightarrow
\ell^- P^+$ as well as its charge conjugate $|\Delta L| = 2$
process $N \rightarrow \ell^+ P^-$ are possible and have the
same width. Hence the factor of 2 associated
with the decay width of this mode in Eq.~(\ref{apptotwid}).
Similarly, the $\Delta L = 0$ and its charge conjugate $|\Delta L|
= 2$ process are possible for the decay modes $N \rightarrow
\ell^- V^+$ and $N \rightarrow \ell^-_1 \ell^+_2 \nu_{\ell_2}$ and
hence have a factor of 2 associated with their width in Eq.~(\ref{apptotwid}). 

\section{Comparison of DM annihilation rates}
\label{appen2}

\begin{figure}[h]
    \centering
    \includegraphics[scale=0.5]{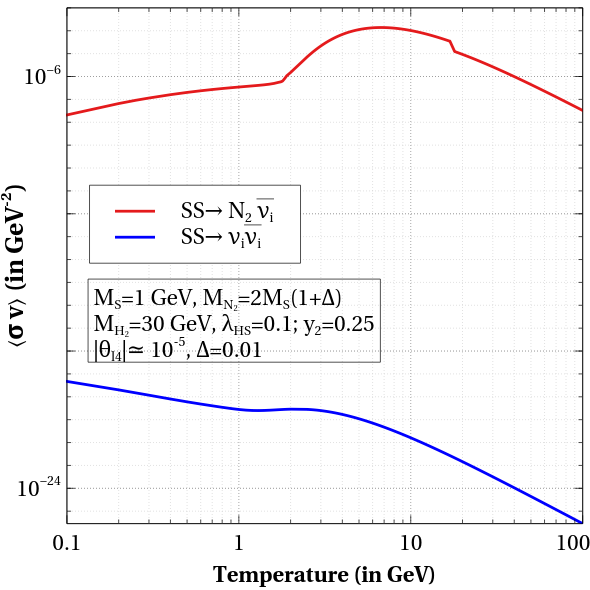}
    \caption{Comparison of the DM annihilation cross-section with an RH neutrino and an active neutrino in the final state (red) and two neutrinos in the final state (blue). The blue line is heavily suppressed due to the active-sterile mixing element, and thus will not contribute to the DM abundances.}
    \label{sigC}
\end{figure}
Light thermal DM in the GeV scale keeps annihilating into different final states even after electroweak symmetry breaking (EWSB). Below EWSB, the SM Higgs acquires a VEV leading to mixing between HNL and SM neutrinos. Therefore, even though DM annihilation into $N \bar{\nu}_\alpha$ remains in the forbidden regime, DM annihilation into $\nu \bar{\nu}$ is always allowed and can occur due to $N-\nu$ mixing. Fig. \ref{sigC} shows the comparison of DM annihilation cross-section to these two final states. As can be seen from the figure, the non-forbidden annihilation into $\nu \bar{\nu}$ final states remains suppressed and negligible compared to that of the forbidden channel. This is due to tiny $N-\nu$ mixing for the chosen $N$ masses required for GeV and sub-GeV scale forbidden DM.

\bibliographystyle{JHEP}
%\bibstyle{apsrev}
\bibliography{references, ref1, ref}

\end{document}